\documentclass[useAMS,usenatbib]{rspublic}

\usepackage{graphicx}
\usepackage{amssymb}
\usepackage{amsmath}
\usepackage{amsfonts}
\usepackage{verbatim}
\usepackage[mathscr]{eucal}

\newcommand{\bfb}{\mbox{\boldmath$b$}}

\newcommand{\bfu}{\mbox{\boldmath$u$}}
\newcommand{\bfU}{\mbox{\boldmath$U$}}
\newcommand{\bfF}{\mbox{\boldmath$F$}}
\newcommand{\bfx}{\mbox{\boldmath$x$}}
\newcommand{\bfzhat}{\mbox{\boldmath $ {\hat z} $ } }
\newcommand{\bfxhat}{\mbox{\boldmath $ {\hat x} $ } }
\newcommand{\bfyhat}{\mbox{\boldmath $ {\hat y} $ } }
\newcommand{\bfB}{\mbox{\boldmath$B$}}

\newcommand{\bfV}{\mbox{\boldmath$V$}}
\newcommand{\bfH}{\mbox{\boldmath$H$}}
\newcommand{\bfv}{\mbox{\boldmath$v$}}
\newcommand{\bfh}{\mbox{\boldmath$h$}}
\newcommand{\bfk}{\mbox{\boldmath$k$}}

\newcommand{\bfalpha}{\boldsymbol{\alpha}}
\newcommand{\bfGamma}{\boldsymbol{\Gamma}}
\newcommand{\bfbeta}{\boldsymbol{\beta}}

\title[Self-consistent mean field MHD]{Self-consistent mean field MHD}
\author[A. Courvoisier, D. W. Hughes, M. R. E. Proctor]{A. Courvoisier$^{1}$\thanks{E-mail:
alice@maths.leeds.ac.uk}, D. W. Hughes$^{1}$, M. R. E. Proctor$^{2}$}

\affiliation{$^1$Department of Applied Mathematics, University of Leeds, Leeds LS2 9JT, UK \\
$^2$D.A.M.T.P., Centre for Mathematical Sciences, University of Cambridge, Cambridge CB3 0WA, UK}

\begin{document}

\date{Accepted . Received ; in original form }


\maketitle

\begin{abstract}
{stellar dynamos, mean field electrodynamics, $\alpha$-effect}
We consider the linear stability of two-dimensional nonlinear magnetohydrodynamic basic states to long-wavelength three-dimensional perturbations. Following Hughes \& Proctor (2009a), the 2D basic states are obtained from a specific forcing function in the presence of an initially uniform mean field of strength $\mathcal{B}$. By extending to the nonlinear regime the kinematic analysis of Roberts (1970), we show that it is possible to predict the growth rate of these perturbations by applying mean field theory to \textit{both} the momentum and the induction equations. If $\mathcal{B}=0$, these equations decouple and large-scale magnetic and velocity perturbations may grow via the kinematic $\alpha$-effect and the AKA instability respectively. However, if $\mathcal{B} \neq 0$, the momentum and induction equations are coupled by the Lorentz force; in this case, we show that four transport tensors are now necessary to determine the growth rate of the perturbations. We illustrate these situations by numerical examples; in particular, we show that a mean field description of the nonlinear regime based solely on a quenched $\alpha$ coefficient is incorrect. 
\end{abstract}


\section{Introduction}

Astronomical observations show that most astrophysical objects are magnetised and often possess a coherent large-scale magnetic field. Such fields are believed to be the results of a \textit{large-scale} dynamo, whereby inductive motions within a conductive fluid can generate and sustain a magnetic field on scales much larger than their own.

The first step in addressing this problem is to understand how a weak seed field can be amplified by the motion. To study this kinematic phase, it is assumed that the magnetic field has no effect on the motion, which can therefore be considered as prescribed. Kinematic large-scale dynamos are traditionally studied within the framework of mean field electrodynamics, a turbulence closure theory in which the effects of small scale interactions are parametrised by transport coefficients. Within this theory, the equation for the evolution of a large-scale magnetic field $\overline{\bfB}$ driven by a motion $\bfU$ (with no mean) is typically of the form
\begin{equation}
\partial_t \overline{\bfB}  = \nabla \times \left( \bfalpha \cdot \overline{\bfB} + \bfbeta \cdot \nabla \overline{\bfB}+ Rm^{-1}\nabla \times \overline{\bfB} \right),
\label{eqn:mf}
\end{equation}
where $\bfalpha$ and $\bfbeta$ are transport pseudo-tensors that depend on the characteristics of $\bfU$ and on the magnetic Reynolds number $Rm$ (see, for example, Moffatt 1978). The symmetric part of $\bfalpha$ is the so-called $\alpha$-effect, which is responsible for the growth of $\overline{\bfB}$. For isotropic turbulence, we have simply that $\alpha_{ij} = \alpha \,\delta_{ij}$ and $\beta_{ijk} = \beta \,\epsilon_{ijk}$; in this case, the growth rate $\sigma$ of a long-wavelength mean field with wavenumber $k \ll 1$ can be approximated by
\begin{equation}
\sigma \approx \alpha k - (\beta + Rm^{-1}) k^2.
\label{eqn:sigma}
\end{equation}
(For completeness, it should be noted, as discussed by Hughes \& Proctor (2009b), that although equations~(\ref{eqn:mf}) and (\ref{eqn:sigma}) are entirely consistent, there is a potential inconsistency between the diffusion term in (\ref{eqn:sigma}) and that derived from considering the decay of a large-scale field. This, however, is not germane to the problem studied in this paper, in which we consider only effects of first order in $k$.)

If $\sigma >0$, an initially weak large-scale magnetic field will grow exponentially until a stage is reached at which the back-reaction of the magnetic field on the flow $\bfU$ must be taken into account. This implies that the induction equation has to be solved in conjunction with the momentum equation, in which the Lorentz forces are included. The hope, however, has been that mean field electrodynamics is still applicable with $\overline{\bfB}$ still satisfying (\ref{eqn:mf}), but with transport coefficients that now depend on the mean field, i.e.\ $\bfalpha = \bfalpha(\overline{\bfB})$ and $\bfbeta = \bfbeta(\overline{\bfB})$. Over the past two decades, considerable effort has been devoted to the analytical and numerical determination of these coefficients and their exact dependence on $\overline{\bfB}$, the so-called $\alpha$-quenching issue (e.g.\ Vainshtein \& Cattaneo 1992, Kulsrud \& Anderson 1992, Gruzinov \& Diamond 1994, Cattaneo \& Hughes 1996, Blackman \& Field 2000; an extensive list of references may be found in the review by Brandenburg \& Subramanian 2005). Recently, it has been suggested that the dependence of $\bfalpha = \bfalpha(\overline{\bfB})$ and $\bfbeta = \bfbeta(\overline{\bfB})$ on $\overline{\bfB}$ can be determined by considering only the effects of the velocity field modified by the Lorentz force (Tilgner \& Brandenburg 2008). In this paper we argue that such an approach, and indeed any approach that privileges the magnetic field over the velocity field in a magnetohydrodynamic (MHD) state, is flawed. We explain how mean field electrodynamics can be applied to the nonlinear regime only if the coupling between the induction and the momentum equation via the Lorentz force is \textit{fully} taken into account. As a result, new transport tensors emerge that are simply not captured by considering the induction equation alone.

Our aim therefore is to consider long-wavelength perturbations to a nonlinear MHD state, and to explain how the growth (or decay) of such perturbations can be explained in terms of transport coefficients arising from small-scale interactions. This can be regarded as the dynamic analogue of the kinematic dynamo problem, in which the growth of a long-wavelength magnetic perturbation to a purely hydrodynamic state can be described in terms of the $\alpha$-tensor. The crucial feature of our analysis of a coupled MHD state is that the velocity and magnetic fields must be treated on an equal footing throughout; failure to do this will lead to physically meaningless results.

In order to make clear the essential features involved, we shall in this paper confine our attention to the restricted problem of investigating 3D long-wavelength linear perturbations to 2D basic MHD states --- the basic state variables depend on $x$, $y$ and $t$, the perturbations are taken to have a long wavelength in the $z$ direction. There are three great advantages in considering this slightly restricted 2D problem. The first is that it is conceptually simpler than considering fully 3D basic states; the second is that average quantities are readily calculated; the third is that it allows a direct comparison between the growth rates determined via evaluation of the various transport tensors and those calculated by a direct solution of the stability problem. Although the generalisation to 3D is formally straightforward (the corresponding equations are derived in Appendix~A), there are a number of subtle issues, not present in 2D, that arise in the measurement and interpretation of the transport tensors. We shall discuss these in a future paper (Courvoisier, Hughes \& Proctor 2010).

We follow the approach recently taken by Hughes \& Proctor (2009a), who considered the growth of linear 3D perturbations of 2D MHD basic states obtained from a specified forcing function in the presence of an imposed initially uniform magnetic field of strength $\mathcal{B}$. Crucially, their analysis involved solving a system of coupled equations for \textit{both} magnetic and velocity perturbations. In this paper, we use exactly the same framework but, this time, we are interested in the stability properties of long-wavelength perturbations. Although these are not in general the dominant modes of instability, they are the modes that can be rigorously described using mean field theory, as discussed above. If the imposed field is small ($\mathcal{B} \ll 1$), the velocity field of the basic state, $\bfU$ say, is determined solely by the prescribed forcing. We then recover the kinematic dynamo problem, for which mean field electrodynamics can be readily applied and in which large-scale magnetic perturbations can grow via the $\alpha$-effect, provided $\bfU$ lacks parity invariance. Furthermore, if $\bfU$ is also anisotropic, large-scale velocity perturbations may grow via the anisotropic kinetic alpha (AKA) instability (Frisch, She \& Sulem 1987). For larger values of $\mathcal{B}$ however, the induction and momentum equations are coupled and cannot be considered separately. For this case, we develop a `nonlinear' theory of mean field electrodynamics. 

The format of the paper is as follows. In \S\,\ref{sec:analysis}, in order to investigate, in general terms, the long-wavelength 3D linear instabilities of 2D nonlinear MHD basic states, we pursue an analysis \textit{\`a la} Roberts (1970), who considered the $\alpha$-effect in the kinematic regime. The main result of our paper --- the derivation of the four transport tensors necessary for the description of long-wavelength perturbations --- is contained in \S\,\ref{sec:analysis}(\ref{sec:full}). In \S\,\ref{sec:analysis}(\ref{sec:Uonly}) we show how, for a purely hydrodynamic basic state (no imposed magnetic field), the evolutions of the magnetic and velocity perturbations decouple, leading to the possibility of a kinematic dynamo (for magnetic perturbations) or the AKA instability (for velocity perturbations). In \S\,\ref{sec:analysis}(\ref{sec:Bonly}) we discuss what might be thought of as the kinematic dynamo problem for a fully coupled MHD flow --- in other words, considering magnetic perturbations but, arbitrarily, neglecting velocity perturbations. Although this problem has received some attention, we argue that such an approach is physically inconsistent. We then present two numerical examples. The first, described in \S\,\ref{sec:numAKA}, illustrates our theory for both hydrodynamic and magnetohydrodynamic basic states. The second, detailed in \S\,\ref{sec:numMWp}, is designed to emphasise that considering the $\alpha$-effect alone fails to describe accurately large-scale dynamo action in the nonlinear regime. Finally, we discuss the implications of our results in \S\,\ref{sec:ccl}.

\section{Analytical approach}\label{sec:analysis}

\subsection{The 2D basic state}

We consider, with the usual notation, the basic state $\bfB(x,y,t)$, $\bfU(x,y,t)$, $P(x,y,t)$ resulting from the solution of the incompressible MHD equations,
\begin{eqnarray}
&\left(\partial_{t} - Rm^{-1} \nabla^{2}  \right) \bfB =  \nabla \times \left( \bfU \times \bfB \right) + \boldsymbol{\mathcal{B}}\cdot \nabla\bfU, \label{eqn:indBS}\\
&\left(\partial_{t} - Re^{-1} \nabla^{2}\right)\bfU = -\nabla P +  \nabla\cdot\left(\bfB\bfB -\bfU\bfU\right) + \boldsymbol{\mathcal{B}} \cdot \nabla \bfB + \bfF, \label{eqn:momBS}\\
&\nabla\cdot\bfB = 0, \quad \nabla\cdot\bfU = 0,\label{eqn:contBS}
\end{eqnarray}
where the forcing $\bfF(x,y,t)$ is spatially periodic and $\boldsymbol{\mathcal{B}}$ is a constant vector field; $Re$ and $Rm$ are, respectively, the kinetic and magnetic Reynolds numbers. Since dynamo action cannot regenerate a 2D magnetic field (Cowling 1933), $\bfB$ is due solely to the distortion of $\boldsymbol{\mathcal{B}}$ by the fluid motion. Furthermore, since $\bfU$ is independent of $z$, the mean flux $\boldsymbol{\mathcal{B}}$ needs to be horizontal in order to have a non-zero $\bfB$ (for convenience, we shall refer to the $xy$-plane as the horizontal plane, although gravity plays no part in the present study).

In the analysis that follows, we consider basic states that are both spatially and temporally periodic, for which it is natural to define an averaging operation by
\begin{equation}
\langle \bfU \rangle = \frac{1}{4\pi^2 T}\int_0^T \int_0^{2\pi}\int_0^{2\pi} \bfU(x,y,t)\, dx\,  dy \, dt,
\label{eqn:av}
\end{equation}
where the flow has spatial periodicity $2 \pi$ and temporal period $T$. We restrict our analysis to the case where $\langle \bfB \rangle = \langle \bfU \rangle = \textbf{0}$, i.e.\ there is no mean flow and the mean magnetic field corresponds to $\boldsymbol{\mathcal{B}}$.

Of course, systems described by (\ref{eqn:indBS}) -- (\ref{eqn:contBS}) can exhibit complex temporal behaviour even with a time-periodic forcing, as shown by Hughes \& Proctor (2009a). In such cases, our analysis is still valid provided that averages are taken over times long compared with the characteristic time scale of the fluctuations in the basic state.

\subsection{Growth rate of $z$-dependent perturbations}\label{sec:full}

We now consider small-scale, incompressible, linear perturbations to the basic state, of the form
\begin{equation}
\left(\bfb(\bfx, t), \bfu(\bfx, t), p(\bfx, t)\right) = \left(\bfH(x,y,t), \bfV(x,y,t),\Pi(x,y,t)\right)\,e^{ikz +p(k)t},
\label{eqn:kansatz}
\end{equation}
where $\bfV$, $\bfH$ and $\Pi$ vary on the same spatial and temporal scales as the basic state, and where $p(k) = \sigma(k)+i\omega(k)$ is the complex growth rate of the perturbation, which depends on its wavenumber $k$. 

The perturbations can be decomposed into average and fluctuating parts; following Roberts (1970) we define the average by
\begin{equation}
\left(\overline{\bfb}(z,t),\overline{\bfu}(z,t) ,\overline{p}(z,t)\right) = \left( \langle\bfH\rangle,\langle\bfV\rangle,\langle\Pi\rangle\right)\,e^{ikz +p(k)t}.
\label{}
\end{equation}
Our aim is to illustrate how a mean field approach can be used to determine the growth rate $p(k)$ of modes with non-zero mean magnetic fields, i.e.\ modes for which $\langle \bfH \rangle \neq \textbf{0}$, in the limit of $k \to 0$.

The equations satisfied by the linear perturbations are given by
\begin{align}
\left(p+Rm^{-1}k^2\right)\bfH + \left(\partial_t - Rm^{-1}\nabla^2 \right) \bfH  &= \left(\nabla + ik\bfzhat \right)\times \left( \bfV \times \bfB + \bfU \times \bfH\right) \nonumber \\
&+ \boldsymbol{\mathcal{B}} \cdot \nabla \bfV ,\label{eqn:indZPERT} \\
\left(p+Re^{-1}k^2\right) \bfV +\left( \partial_t - Re^{-1}\nabla^2 \right) \bfV &= \left(\nabla + ik\bfzhat \right)\cdot \left( \bfH\bfB + \bfB\bfH -\bfV\bfU - \bfU\bfV\right) \nonumber \\
 &-\nabla\Pi- ik\Pi\bfzhat  + \boldsymbol{\mathcal{B}} \cdot \nabla \bfH ,\label{eqn:momZPERT} \\
\nabla \cdot \bfH + i k\bfzhat \cdot \bfH = 0, \quad
&\nabla \cdot \bfV + i k\bfzhat \cdot \bfV = 0. \label{eqn:contZPERT}
\end{align}
We now treat $k$ as a small parameter and expand $\bfH$, $\bfV$, $\Pi$ and the growth rate $p$ in powers of $k$, so that, for example,
\begin{equation}
\bfH = \bfH_0+\bfH_1+ \ldots \bfH_n +\ldots,
\label{}
\end{equation}
where $\bfH_n$ is of $n^{th}$ order in $k$. Each $\bfH_n$ can be decomposed into its mean and fluctuating parts, $\bfH_n = \langle \bfH_n \rangle + \bfh_n$; similarly, $\bfV_n = \langle \bfV_n \rangle + \bfv_n$.

At zeroth order, equations~(\ref{eqn:indZPERT}) -- (\ref{eqn:contZPERT}) become
\begin{align}
p_0\bfH_0 + \left(\partial_t - Rm^{-1}\nabla^2 \right) \bfH_0  &= \nabla\times \left( \bfV_0 \times \bfB + \bfU \times \bfH_0\right) + \boldsymbol{\mathcal{B}} \cdot \nabla \bfV_0,
\label{eqn:indZERO} \\
p_0 \bfV_0 +\left( \partial_t - Re^{-1}\nabla^2 \right) \bfV_0 &=\nabla \cdot \left( \bfH_0\bfB + \bfB\bfH_0 -\bfV_0\bfU - \bfU\bfV_0\right) \nonumber \\
&-\nabla\Pi_0 + \boldsymbol{\mathcal{B}} \cdot \nabla \bfH_0,
\label{eqn:momZERO} \\
\nabla \cdot \bfH_0 = 0, \quad
&\nabla \cdot \bfV_0 = 0. \label{eqn:contZERO}
\end{align}
On averaging these equations, we obtain
\begin{equation}
p_0 \langle \bfH_0 \rangle = p_0\langle \bfV_0 \rangle = 0.
\label{}
\end{equation}
For non-zero mean field solutions it follows therefore that $p_0 = 0$. The corresponding equations for the fluctuations then become
\begin{align}
(\partial_t - Rm^{-1}\nabla^2 ) \bfh_0  &= 
\nabla\times \left( \bfv_0 \times \bfB + \bfU \times \bfh_0\right)\nonumber \\ &+ \langle \bfH_0 \rangle \cdot \nabla\bfU -\langle \bfV_0 \rangle\cdot\nabla\bfB + \boldsymbol{\mathcal{B}} \cdot \nabla \bfv_0,
\label{eqn:indZEROFLUCT} \\
( \partial_t - Re^{-1}\nabla^2 ) \bfv_0 &=-\nabla\Pi_0 + \nabla \cdot \left( \bfh_0\bfB + \bfB\bfh_0 -\bfv_0\bfU - \bfU\bfv_0\right)\nonumber \\
&+ \langle \bfH_0 \rangle \cdot \nabla\bfB -\langle \bfV_0 \rangle\cdot\nabla\bfU + \boldsymbol{\mathcal{B}} \cdot \nabla \bfh_0,
\label{eqn:momZEROFLUCT} \\
\nabla \cdot \bfh_0 = 0, \quad
&\nabla \cdot \bfv_0 = 0. \label{eqn:contZEROFLUCT}
\end{align}
The first non-trivial mean field equations are obtained from the horizontal averages of equations~(\ref{eqn:indZPERT}) -- (\ref{eqn:contZPERT}) at order $k$; we obtain
\begin{align}
&p_1 \langle \bfH_0 \rangle = ik\bfzhat\times \langle \bfv_0 \times \bfB + \bfU \times \bfh_0 \rangle, \label{eqn:1stavB} \\
&p_1 \langle \bfV_0 \rangle = -i \Pi_0k\,\bfzhat +ik\bfzhat\cdot \langle \bfh_0 \bfB + \bfB\bfh_0 - \bfv_0\bfU -\bfU\bfv_0 \rangle, \label{eqn:1stavU} \\
&\bfzhat \cdot \langle\bfH_0 \rangle= 0, \quad
\bfzhat \cdot \langle\bfV_0 \rangle= 0. \label{eqn:1stavdivhv} 
\end{align}
It can be seen from equations~(\ref{eqn:indZEROFLUCT}) -- (\ref{eqn:contZEROFLUCT}) that $\bfh_0$ and $\bfv_0$ are subject to a linear forcing by both $\langle \bfH_0 \rangle$ and $\langle \bfV_0 \rangle$. Therefore,~(\ref{eqn:1stavB}) and~(\ref{eqn:1stavU}) can be rewritten (without approximation) as
\begin{align}
&p_1 \langle \bfH_0 \rangle = ik\bfzhat\times \left( \bfalpha^{B}\cdot\langle \bfH_0 \rangle + \bfalpha^{U}\cdot\langle \bfV_0 \rangle \right), \label{eqn:1stavB2} \\
&p_1 \langle \bfV_0 \rangle = -ik\bfzhat \Pi_0 +ik\bfzhat\cdot \left( \bfGamma^{B}\cdot\langle \bfH_0 \rangle + \bfGamma^{U}\cdot\langle \bfV_0 \rangle \right), \label{eqn:1stavU2} 
\end{align}
or, in component form, as
\begin{align}
&p_1 \langle H_0 \rangle_\ell = -i k \epsilon_{3 \ell m} \left( \alpha^B_{mn} \langle H_0 \rangle_n + \alpha^U_{mn} \langle V_0 \rangle_n \right), 
\label{eqn:1stavB2a} \\
&p_1 \langle V_0 \rangle_\ell = -i k \delta_{\ell 3} \Pi_0 + i k \left( \Gamma^B_{3 \ell m} \langle H_0 \rangle_m  
+ \bfGamma^U_{3 \ell m} \langle V_0 \rangle_m  \right) . 
\label{eqn:1stavU2a} 
\end{align}
The tensors\footnote{To be more precise, $\bfalpha^{U}$ and $\bfGamma^{U}$ are true tensors, whereas $\bfalpha^{B}$ and $\bfGamma^{B}$ are pseudo-tensors.}  $\bfalpha^{B}$, $\bfalpha^{U}$, $\bfGamma^{B}$ and $\bfGamma^{U}$ are constant and depend on the basic state $\bfU$ and $\bfB$ --- hence on the forcing $\bfF$, the imposed magnetic field $\boldsymbol{\mathcal{B}}$ and the Reynolds numbers. The tensors $\bfGamma^{U}$ and $\bfGamma^{B}$ are symmetric with respect to their first two indices.

In order to use mean field theory to determine the growth rate of long-wavelength perturbations to an MHD state, it is necessary to determine the four tensors $\bfalpha^{U}$, $\bfalpha^{B}$, $\bfGamma^{U}$ and $\bfGamma^{B}$ by solving equations~(\ref{eqn:indZEROFLUCT}) -- (\ref{eqn:contZEROFLUCT}) for $\bfh_0$ and $\bfv_0$, with imposed constant magnetic and velocity fields, alongside the basic state equations~(\ref{eqn:indBS}) -- (\ref{eqn:contBS}). 

We note that for the 2D system considered here, the generated mean fields are horizontal. Furthermore, only the horizontal component of the mean emf, $\langle \bfv_0 \times \bfB + \bfU \times \bfh_0 \rangle$, contributes to the growth of the mean magnetic field; therefore only the $2 \times 2$ parts of $\bfalpha^{U}$ and $\bfalpha^{B}$ relating horizontal quantities need be calculated. Similarly, only the horizontal component of the right hand side of (\ref{eqn:1stavU}) influences $p_1$, so the only components of the mean stress tensor of interest are of the form $\langle h_{03}\,B_\ell + B_3 \, h_{0 \ell} - v_{03}\,U_\ell - U_3 \, v_{0 \ell} \rangle$, with $\ell = 1,2$. Thus we need only to determine $\Gamma^U_{3 \ell m} = \Gamma^U_{\ell 3m}$ and $\Gamma^B_{3 \ell m} = \Gamma^B_{\ell 3m}$ for $\ell, m=1,2$.

Equations~(\ref{eqn:1stavB2}) and~(\ref{eqn:1stavU2}) can then be rewritten as
\begin{equation}
p_1 \left[ \begin{array}{c} H_{0x} \\ H_{0y} \\ V_{0x} \\ V_{0y} \end{array}\right] = ik
\left[ \begin{array}{cccc} 
-\alpha_{21}^B & -\alpha_{22}^B & -\alpha_{21}^U & -\alpha_{22}^U \\ 
\quad\alpha_{11}^B & \quad\alpha_{12}^B & \quad\alpha_{11}^U & \quad\alpha_{12}^U \\ 
\Gamma_{131}^B & \Gamma_{132}^B & \Gamma_{131}^U & \Gamma_{132}^U \\ 
\Gamma_{231}^B & \Gamma_{232}^B & \Gamma_{231}^U & \Gamma_{232}^U 
\end{array}\right]
\left[ \begin{array}{c} H_{0x} \\ H_{0y} \\ V_{0x} \\ V_{0y} \end{array}\right] = \boldsymbol{\mathcal{M}} \, k\, \left[ \begin{array}{c} H_{0x} \\ H_{0y} \\ V_{0x} \\ V_{0y} \end{array}\right] .
\label{}
\end{equation}
The first order growth rate is then an eigenvalue of the matrix $\boldsymbol{\mathcal{M}}\, k$.

In \S\S\,\ref{sec:numAKA} and~\ref{sec:numMWp}, we test our analysis numerically by comparing the actual growth rate of $z$-dependent perturbations with predictions obtained by calculating the coefficients of the four tensors $\bfalpha^{U}$, $\bfalpha^{B}$, $\bfGamma^{U}$ and $\bfGamma^{B}$. Prior to this though, we consider two special cases of the above theory.

\subsection{Purely hydrodynamic basic states, $\boldsymbol{\mathcal{B}}=\textbf{0}$}\label{sec:Uonly}

In this subsection, we consider the limit when $\boldsymbol{\mathcal{B}}=\textbf{0}$. This implies that $\bfB=\textbf{0}$ and the basic state is purely hydrodynamic. 
In this case, equations (\ref{eqn:indZEROFLUCT}) and (\ref{eqn:momZEROFLUCT}) for the magnetic and velocity perturbations simply become
\begin{align}
(\partial_t - Rm^{-1}\nabla^2 ) \bfh_0  &= \nabla\times \left( \bfU \times \bfh_0\right) + \langle \bfH_0 \rangle \cdot \nabla\bfU ,
\label{eqn:indnoB} \\
( \partial_t - Re^{-1}\nabla^2 ) \bfv_0 &=-\nabla\Pi_0 - \nabla \cdot \left(\bfv_0\bfU + \bfU\bfv_0\right) -\langle \bfV_0 \rangle\cdot\nabla\bfU.
\label{eqn:momnoB} 
\end{align}
The corresponding first order equations giving the growth rate, (\ref{eqn:1stavB}) and (\ref{eqn:1stavU}), become
\begin{align}
&p_1^B \langle \bfH_0 \rangle = ik\bfzhat\times \langle  \bfU \times \bfh_0 \rangle = ik\bfzhat\times \left( \bfalpha \cdot \langle \bfH_0 \rangle \right), \label{eqn:1stavB4} \\
&p_1^U \langle \bfV_0 \rangle = -i \Pi_0k\,\bfzhat -ik\bfzhat\cdot \langle  \bfv_0\bfU +\bfU\bfv_0 \rangle = ik\bfzhat\cdot \left( \bfGamma \cdot\langle \bfV_0 \rangle \right) . \label{eqn:1stavU4}  
\end{align}

We see that in this limit the evolutions of $\bfH_0$ and $\bfV_0$ decouple, and so we expect the growth rates $p_1^B$ and $p_1^U$ to be distinct. The magnetic problem, described by (\ref{eqn:indnoB}) and (\ref{eqn:1stavB4}), reduces to the kinematic dynamo problem for the velocity field $\bfU$, with $\bfalpha$ the standard $\bfalpha$ tensor of mean-field electrodynamics. The symmetric part of this (pseudo)-tensor encompasses the $\alpha$-effect of kinematic dynamo theory; it can be non-zero only in flows lacking parity invariance, such as helical flows (see Moffatt 1978). The hydrodynamic problem, described by (\ref{eqn:momnoB}) and (\ref{eqn:1stavU4}), reduces to that studied by Frisch \textit{et al.}\ (1987). The coefficient $\bfGamma$ here corresponds to the anisotropic kinetic alpha (AKA) effect, which can be nonzero in anisotropic flows lacking parity invariance. This effect also requires that a uniform large-scale velocity field produces a change in the average small-scale Reynolds stresses, therefore that Galilean invariance be broken. For more details and a comparison between the conditions required for the existence of the $\alpha$ and AKA effects, see Frisch \textit{et al.}\ (1987).

Equations~(\ref{eqn:1stavB4}) and~(\ref{eqn:1stavU4}) can be rewritten in matrix form as
\begin{align}
p^B_1 \left[ \begin{array}{c} H_{0x} \\ H_{0y} \end{array}\right] = ik
\left[ \begin{array}{cc} 
-\alpha_{21} & -\alpha_{22}  \\ 
\quad\alpha_{11} & \quad\alpha_{12}   
\end{array}\right]
\left[ \begin{array}{c} H_{0x} \\ H_{0y} \end{array}\right] = \boldsymbol{\mathcal{A}} \, k\,\left[ \begin{array}{c} H_{0x} \\ H_{0y} \end{array}\right], \\
p^U_1 \left[ \begin{array}{c} V_{0x} \\ V_{0y} \end{array}\right] = ik
\left[ \begin{array}{cc} 
\quad\Gamma_{131} & \quad\Gamma_{132} \\ 
\quad\Gamma_{231} & \quad\Gamma_{232} 
\end{array}\right]
\left[ \begin{array}{c} V_{0x} \\ V_{0y} \end{array}\right] = \boldsymbol{\mathcal{G}} \, k\, \left[ \begin{array}{c}V_{0x} \\ V_{0y} \end{array}\right],
\label{}
\end{align}
so the growth rates $p_1^B$ and $p_1^U$ are eigenvalues of the matrices $\boldsymbol{\mathcal{A}}\, k$ and $\boldsymbol{\mathcal{G}}\, k$, respectively.
The $\alpha$-effect can be readily determined by solving~(\ref{eqn:indnoB}) for an imposed horizontal mean field $\langle \bfH_0 \rangle$, while $\bfGamma$ can be obtained by solving~(\ref{eqn:momnoB}) for an imposed horizontal mean flow $\langle \bfV_0 \rangle$.

The limiting procedure under consideration here (i.e.\ taking $\boldsymbol{\mathcal{B}}=\textbf{0}$) constitutes a well-defined physical problem. It corresponds to the case when there is no Lorentz force (i.e.\ in 2D there is no mean flux, or, in 3D, $Rm$ and $Re$ are low enough to prevent small-scale dynamo action and small-scale hydrodynamic instabilities); the momentum and induction equations therefore decouple and the $\alpha$ and AKA instabilities can exist independently. A numerical example of a forcing for which both $\alpha$ and AKA-effects are present is given in \S\,\ref{sec:numAKA}, in which we also describe what happens when $\boldsymbol{\mathcal{B}} \neq \textbf{0}$.

\subsection{Magnetic perturbations only, $\bfu = \textbf{0}$}\label{sec:Bonly}

In this subsection, we consider the problem of perturbing the basic state only in the magnetic field, i.e.\ we arbitrarily set $\bfu = \textbf{0}$, and hence $\bfV = \textbf{0}$. This corresponds to studying a \textit{kinematic} dynamo problem for the basic state velocity field $\bfU(\mathcal{B})$. This line of research has been considered recently by Cattaneo \& Tobias (2009) and by Tilgner \& Brandenburg (2008).

In this case, equation (\ref{eqn:indZEROFLUCT}) for the magnetic perturbation at zeroth order simply becomes
\begin{equation}
\left(\partial_{t} - Rm^{-1} \nabla^{2}  \right) \bfh_0 =  \nabla \times \left( \bfU(\mathcal{B}) \times \bfh_0 \right) + \langle \bfH_0 \rangle \cdot \nabla\bfU(\mathcal{B}).
\label{eqn:indPERTB}
\end{equation}
Our notation here, which shows that the basic state flow $\bfU$, and hence $\bfh_0$, depends on $\boldsymbol{\mathcal{B}}$, is chosen to emphases the difference between~(\ref{eqn:indnoB}) and~(\ref{eqn:indPERTB})

Equations (\ref{eqn:1stavB}) and (\ref{eqn:1stavB2}), which give the growth rate, become
\begin{equation}
p_1^B \langle \bfH_0 \rangle = ik\bfzhat\times \langle \bfU(\mathcal{B}) \times \bfh_0 \rangle = ik\bfzhat\times \left( \bfalpha(\mathcal{B}) \cdot \langle \bfH_0 \rangle \right),
\label{eqn:1stavB3}
\end{equation}
where $\bfalpha(\mathcal{B})$ corresponds to the $\bfalpha$ tensor of mean field electrodynamics. It depends on the mean flux of the basic state, $\mathcal{B}$, as well as on $\bfF$ and the Reynolds numbers; in that sense, it is distinct from the purely kinematic $\bfalpha$ tensor described in \S\,\ref{sec:analysis}(\ref{sec:Uonly}). We expect $\bfalpha(\mathcal{B})$ to be different from $\bfalpha^B$ in general, since the latter encompasses the dependence of \textit{both} $\bfh_0$ and $\bfv_0$ on the mean magnetic field $\langle \bfH_0 \rangle$. 

Using (\ref{eqn:1stavB3}) we can define the matrix $\boldsymbol{\mathcal{A}}(\mathcal{B})$ whose eigenvalues give the growth rate $p_1^B$. Its entries are similar to those of $\boldsymbol{\mathcal{A}}$, but again we emphases the $\mathcal{B}$ dependence in the notation.

The $\bfalpha(\mathcal{B})$ tensor can be readily determined by solving equation~(\ref{eqn:indPERTB}). We note that this equation is formally equivalent to the induction equation~(\ref{eqn:indBS}) from the basic state, with $\bfh_0 \equiv \bfB$ and $\langle \bfH_0 \rangle \equiv \boldsymbol{\mathcal{B}}$; therefore some components of $\bfalpha(\mathcal{B})$ can be directly computed from the mean emf $\langle \bfU(\mathcal{B}) \times \bfB \rangle$ emerging from the basic state. To obtain the other components of $\alpha$, it is necessary to solve a `passive' induction equation alongside the basic state, with the imposed mean field in a direction perpendicular to that of $\boldsymbol{\mathcal{B}}$, i.e. in the present context, this corresponds to solving equation~(\ref{eqn:indPERTB}) with $\langle \bfH_0 \rangle = \langle H_0 \rangle \, \bfyhat$. This procedure for calculating $\bfalpha(\mathcal{B})$ is now often referred to as the `test-field' method (Schrinner \textit{et al.} 2005) and has been discussed in the context of $\alpha$-quenching by, for example, Brandenburg \textit{et al.} (2008), Tilgner \& Brandenburg  (2008).

It is important to note that we can use the basic state to determine the kinematic $\alpha$-effect for $\bfU(\mathcal{B})$ only because the system is two-dimensional. Indeed, equation~(\ref{eqn:indBS}) is part of a system of coupled equations, while~(\ref{eqn:indPERTB}) is a passive vector equation for the same velocity field. The 3D simulations of Cattaneo \& Tobias (2009) show that, despite being identical, these equations in general have different magnetic field solutions. In 2D however, we find that both equations converge to the same solution, owing to the fact that dynamo action is not possible in this case and that therefore the solution is driven solely by the imposed mean magnetic field.

Finally, we would like to stress the important point that considering only magnetic perturbations constitutes a problem with no physical meaning, since, as discussed in the introduction, it privileges the magnetic field over the velocity field in a fully coupled MHD state. Crucially, there is no physical justification for setting $\bfV = \textbf{0}$. Indeed, from  (\ref{eqn:momZEROFLUCT}), this would imply that $\nabla \cdot ( \bfh_0\bfB + \bfB\bfh_0 )+ \boldsymbol{\mathcal{B}} \cdot \nabla \bfh_0 = \textbf{0}$, which, in general, is in contradiction with the assumed existence of non-zero magnetic perturbations. 

As argued by Hughes \& Proctor (2009a), it is therefore necessary to consider perturbations to \textit{both} the induction and momentum equations. In this case, the actual growth rate of the fluctuations is completely unrelated to the kinematic $\alpha$-effect corresponding to $\bfU(\mathcal{B})$, as we shall demonstrate numerically in \S\,\ref{sec:numMWp}. Instead, all four tensors $\bfalpha^{U}$, $\bfalpha^{B}$, $\bfGamma^{U}$ and $\bfGamma^{B}$ need to be determined.

For completeness it should perhaps be noted that there is a related unphysical problem, which would consist of investigating the stability of velocity perturbations to an MHD state, disregarding magnetic perturbations. This would involve the solution of (\ref{eqn:momZEROFLUCT}) but with $\bfh_0 = \bfH_0 =0$, i.e.\
\begin{equation}
( \partial_t - Re^{-1}\nabla^2 ) \bfv_0 = -\nabla \Pi_0 - 
\nabla \cdot \left( \bfv_0 \bfU (\mathcal{B}) + \bfU (\mathcal{B}) \bfv_0 \right) -
\langle \bfV_0 \rangle \cdot \nabla \bfU (\mathcal{B}).
\label{eqn:indPERTU}
\end{equation}

\section{Numerical example I: AKA forcing}\label{sec:numAKA}

\subsection{Basic state and mean field coefficients}

For this example, we choose the force in the momentum equation~(\ref{eqn:momBS}) to be
\begin{equation}
\bfF = \left(Re^{-1}\sqrt{2}\cos{(y+Re^{-1}t)},Re^{-1}\sqrt{2}\cos{(x-Re^{-1}t)},F_x+F_y\right).
\label{eqn:AKAforce}
\end{equation}
This forcing was first proposed by Frisch \textit{et al.}\ (1987) in order to drive a flow displaying the AKA-effect. By design, flows driven by this forcing lack parity invariance, and thus, in general, may be expected to drive both $\alpha$ and AKA effects. Frisch \textit{et al.}\ (1987) determined the $\bfGamma$ tensor analytically when $Re \ll 1$, thus allowing the neglect of the inertia terms from the momentum equation. Here we drive the flow with $Re=1$; in this case, and in contrast to when $Re \ll 1$, the helicity of the ensuing flow is non-zero. We first consider the case when $\mathcal{B}=0$, described in \S\,\ref{sec:analysis}(\ref{sec:Uonly}), and restrict our simulations to $Rm=16$. In this parameter regime, both $\bfU$ and $\bfB$ are time-periodic with a well-defined frequency.
The matrices $\boldsymbol{\mathcal{A}}$ and $\boldsymbol{\mathcal{G}}$ are determined by solving~(\ref{eqn:indnoB}) and~(\ref{eqn:momnoB}) respectively for solenoidal vector fields, which is equivalent to solving the full system~(\ref{eqn:indZEROFLUCT}) --~(\ref{eqn:contZEROFLUCT}), since $\mathcal{B}=0$. The matrices are given in~\ref{app:matrices} and their eigenvalues are contained in table~\ref{tab:evAKA}. These are real and show that the velocity field considered does indeed display both $\alpha$ and AKA effects; the latter is of greater magnitude, which is maybe not unexpected for a velocity field specifically designed to drive an AKA effect and with little kinetic helicity, which, although, not strictly necessary for an $\alpha$-effect, is often helpful (Gilbert, Frisch \& Pouquet 1988).

We then go on to consider the case when $\boldsymbol{\mathcal{B}}=0.1\, \bfxhat$, which leads to the coupling of magnetic and velocity perturbations via the Lorentz force. We determine the matrix $\boldsymbol{\mathcal{M}}$, the elements of which are given in~\ref{app:matrices}, and its eigenvalues, which are given in table~\ref{tab:evAKA}; they show that the growing modes are now oscillatory.

\begin{table}
\begin{center}
\begin{tabular}{|c|c|c|}
\hline
$\mathcal{B}$ & relevant matrix & eigenvalues \\ \hline 
$0$ & $\boldsymbol{\mathcal{A}}$ & $\pm 0.077$  \\
 & $\boldsymbol{\mathcal{G}}$ & $\pm 0.386$  \\ \hline
$0.1$ & $\boldsymbol{\mathcal{M}}$ & $\pm 0.209 + 0.182\, i \qquad \pm 0.171 - 0.176\, i$ \\ \hline
\end{tabular}
\end{center}
\caption{Eigenvalues of the matrices for the AKA forcing; $Re=1$, $Rm=16$.}\label{tab:evAKA}
\end{table}

\subsection{Growth of $z$-dependent perturbations}

For comparison with the mean field theory results, we calculate directly the growth rate of $z$-dependent linear perturbations to the basic state. Following the method of Hughes \& Proctor (2009a), we solve equations~(\ref{eqn:indZPERT}) -- (\ref{eqn:contZPERT}) numerically, using a time stepping procedure, in order to determine $\bfH(x,y,t)\, e^{p(k)t}$ and $\bfV(x,y,t) \,e^{p(k)t}$. We obtain $\sigma(k) = Re[p(k)]$ from a least squares fit to the time series for the corresponding magnetic and kinetic energies, and determine the frequency $\omega(k)=Im[p(k)]$ by inspection of the time evolution of space-averaged magnetic or velocity components. Since we are interested in verifying our predictions for $p(k)$ at low $k$, we only consider values of $k$ less than $0.1$.

\begin{figure}
\begin{center}
\includegraphics[scale = 0.42]{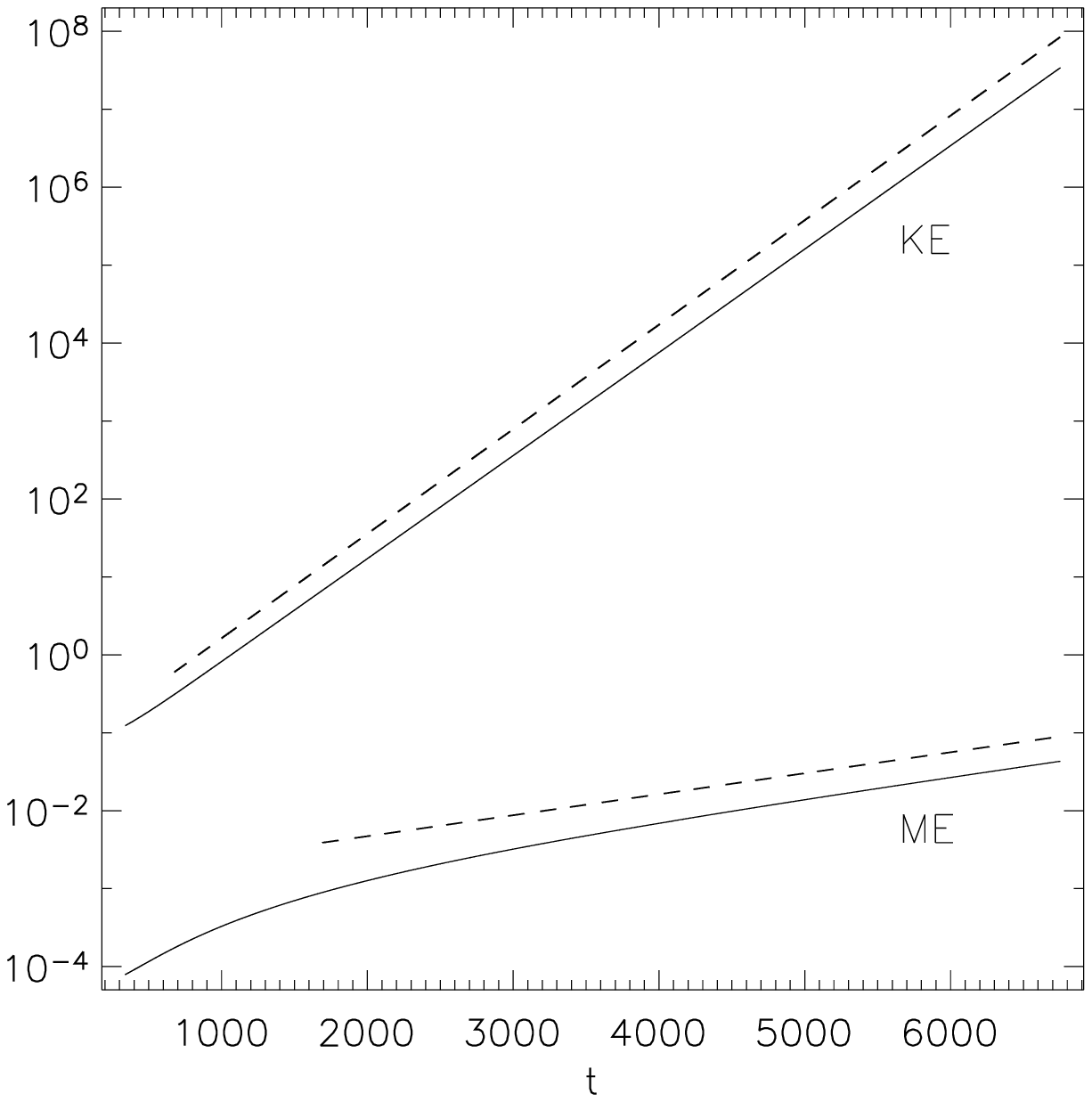}
\includegraphics[scale = 0.42]{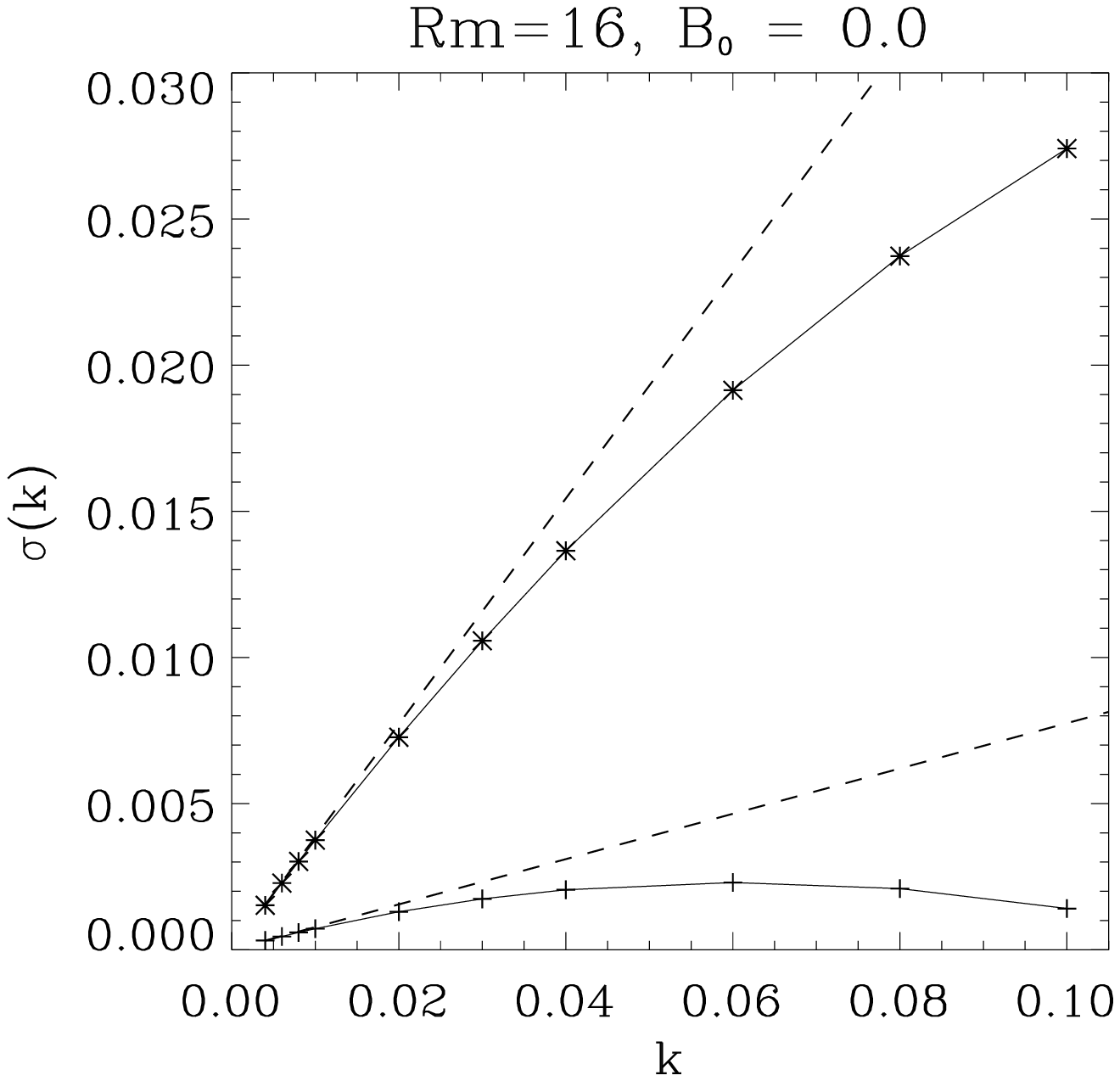}
\end{center}
\caption{\label{fig:AKAB0} 
Results for the AKA forcing; $\mathcal{B}=0$, $Re=1$, $Rm=16$. 
Left: Time series of the energy in the perturbed magnetic and velocity fields for $k=0.004$. The dashed lines show the predicted growth rate for the $\alpha$ and AKA instabilities.
Right: $\sigma(k)$ for the magnetic ($+$) and velocity ($\star$) perturbations. The dashed lines correspond to $\lambda\, k$, where $\lambda$ stands for the positive eigenvalue of $\boldsymbol{\mathcal{A}}$ (for the magnetic perturbations) or of $\boldsymbol{\mathcal{G}}$ (for the velocity perturbations).}
\end{figure}

The results for $\boldsymbol{\mathcal{B}}=0$ are given in figure~\ref{fig:AKAB0}. As discussed above, the induction and momentum equations decouple, thus allowing magnetic and velocity perturbations to evolve independently. The left panel of figure~\ref{fig:AKAB0}, which presents typical time series of the energy in the perturbed magnetic and velocity fields (here for $k=0.004$), confirms that these modes have different growth rates. The right panel summarises the results obtained for $\sigma(k)$ and shows that, for low values of $k$, $\sigma \sim \lambda \, k$, where $\lambda$ stands for the positive eigenvalue of $\boldsymbol{\mathcal{A}}$ or $\boldsymbol{\mathcal{G}}$, for magnetic or velocity perturbations respectively, in agreement with the theory derived in \S\,\ref{sec:analysis}(\ref{sec:Uonly}). 

\begin{figure}
\begin{center}
\includegraphics[scale = 0.42]{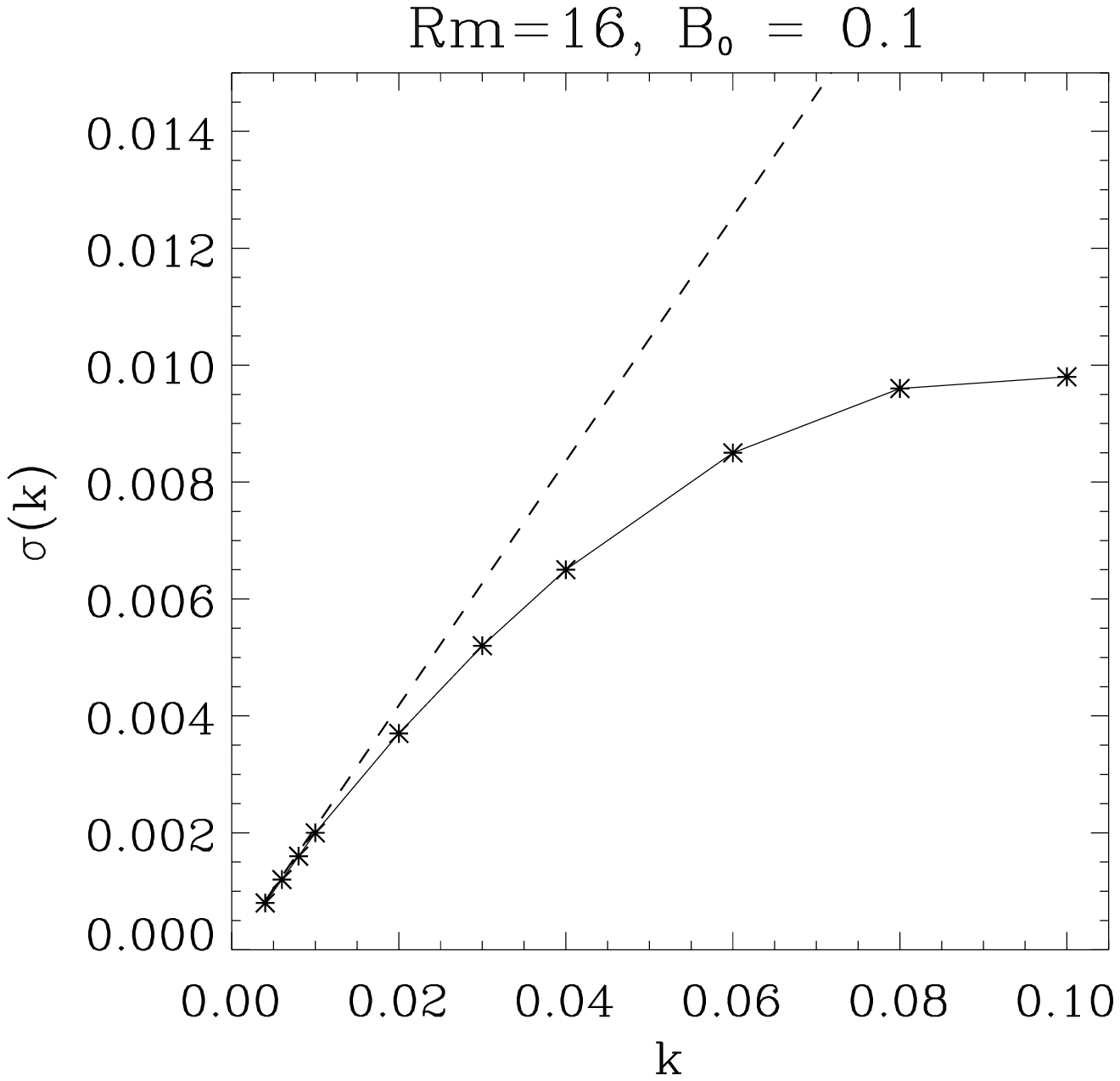}
\includegraphics[scale = 0.42]{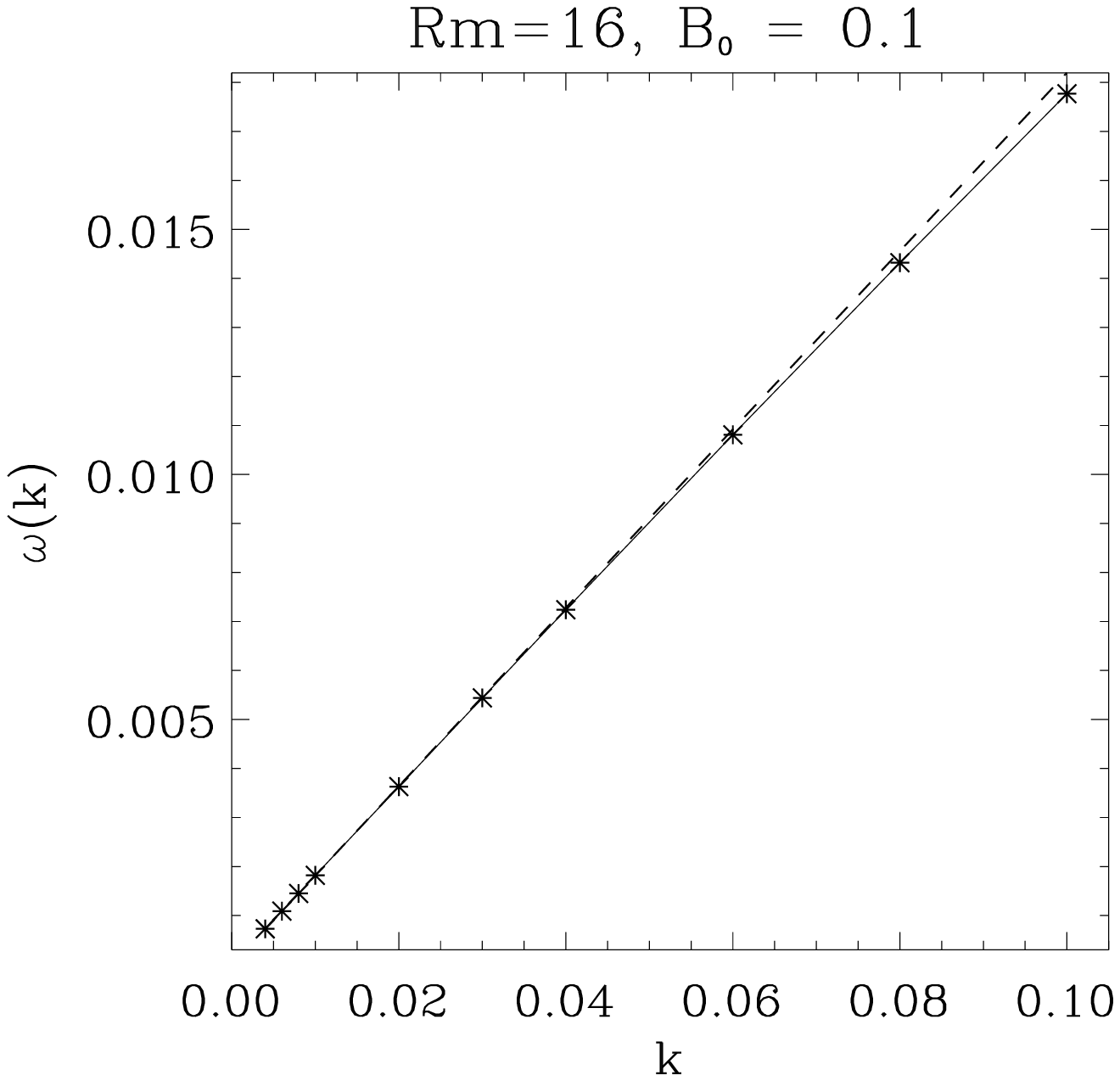}
\end{center}
\caption{\label{fig:AKAB1} 
$\sigma(k)$ (left) and $\omega(k)$ (right) for the AKA forcing; $Rm = 16$, $\mathcal{B}=0.1$. The dashed lines corresponds to $Re[\lambda]\, k$ (left) and $Im[\lambda]\, k$ (right), where $\lambda$ is the eigenvalue of $\boldsymbol{\mathcal{M}}$ with the largest real part.}
\end{figure}

The results for $\mathcal{B}=0.1$ are given in figure~\ref{fig:AKAB1}, which presents $\sigma(k)$ (left panel) and $\omega(k)$ (right panel) together with $Re[\lambda]\, k$ and $Im[\lambda]\, k$ (dashed lines), where $\lambda$ is the eigenvalue of $\boldsymbol{\mathcal{M}}$ with the largest real part. Here again we see that the linear behaviour of both $\sigma(k)$ and $\omega(k)$ at low $k$ is well predicted by the analysis.

Finally, we note that the dynamo problem, including the nonlinear saturation of the instability, in a flow driven by the AKA forcing (or similar) has been considered by, for example, Galanti, Sulem \& Gilbert (1991) and Brandenburg \& von Rekowski (2001). In our notation, this would correspond to studying the saturation of the instability in the case when $\mathcal{B}=0$, i.e. when the basic state is purely hydrodynamic. The focus of the present paper is however the \textit{linear} phase of the instability when $\mathcal{B} \neq 0$.

\section{Numerical example II: MW$+$ forcing}\label{sec:numMWp}

\subsection{The basic state}

We now choose $\textbf{F}$ so that, in the absence of $\boldsymbol{\mathcal{B}}$, it drives the MW$+$ flow of Otani (1993), given by
\begin{equation}
\bfU_0(x,y,t) = (\partial_y \psi, -\partial_x \psi, -\psi) ,
\label{eqn:MWp}
\end{equation}
with
\begin{equation}
\psi(x,y,t) = \cos{x}\cos^2{t}-\cos{y}\sin^2{t}.
\label{eqn:psiMWp}
\end{equation}
This flow has the Beltrami property that $\bfU$ is parallel to $\nabla \times \bfU$ and so $\bfF$ is simply the solution of
\begin{equation}
\bfF = \partial_t \bfU_0 - Re^{-1} \nabla^2 \bfU_0.
\label{}
\end{equation}
To ensure that the flow driven when $\boldsymbol{\mathcal{B}} = \textbf{0}$ settles down rapidly to $\bfU_0$ and to avoid any hydrodynamic instabilities we restrict our computations to $Re=1$.
The symmetry properties of the MW$+$ flow imply that there is no AKA effect and that its kinematic $\bfalpha$ tensor is diagonal, with entries $\alpha_{11} = \alpha_{22} = \alpha$; these have been computed numerically by Courvoisier (2008).

We use this example to illustrate the fact that taking only magnetic perturbations into account can be misleading. To this end, we consider both the full problem and the artificial case of taking $\bfV = \textbf{0}$, discussed in \S\,\ref{sec:analysis}(\ref{sec:Bonly}). As in the previous section, we take $\boldsymbol{\mathcal{B}}$ in the $x$-direction. For the parameters considered here, the basic state proves to be time-periodic with a well-defined frequency.

\begin{table}
\begin{center}
\begin{tabular}{|c|c|c|}
\hline
$Rm$ & $\mathcal{B}$ & eigenvalues \\ \hline 
$16$ & $0.001$ &  $\pm 0.988$  \\ \hline
 & $0.25$  & $\pm 0.624 $\\ \hline
 & $1.0$ & $\pm 0.088 $ \\ \hline
$128$ & $0.1$ & $\pm 1.110 $ \\ \hline
\end{tabular}
\end{center}
\caption{Eigenvalues of $\boldsymbol{\mathcal{A}}(\mathcal{B})$ for the MW$+$ forcing; $Re=1$.}\label{tab:evMWpA}
\end{table}

We first compute the eigenvalues of $\boldsymbol{\mathcal{A}}(\mathcal{B})$ by using the basic state and solving~(\ref{eqn:indPERTB}) with an imposed mean field in the direction perpendicular to $\boldsymbol{\mathcal{B}}$. The results are given in table~\ref{tab:evMWpA} (see~\ref{app:matrices} for the full matrices).
For $\mathcal{B} \ll 1$, $\boldsymbol{\mathcal{A}}(\mathcal{B}) \sim\boldsymbol{\mathcal{A}}$, and we recover the kinematic $\alpha$-effect for the MW$+$ flow. As $\mathcal{B}$ increases for $Rm=16$, we see that the eigenvalues of $\boldsymbol{\mathcal{A}}(\mathcal{B})$ decrease in magnitude, as expected from $\alpha$-quenching studies. This tells us that the flow $\bfU(\mathcal{B})$, taken in isolation, is not as good a \textit{kinematic} large-scale dynamo as the flow $\bfU$.

\begin{table}
\begin{center}
\begin{tabular}{|c|c|c|c|}
\hline
$Rm$ & $\mathcal{B}$ &  eigenvalues & eigenvectors\\ \hline 
$16$ & $0.001$ &  $\pm 0.988$  & $\langle \bfH_0 \rangle$ only\\
&  &  $0$ & \\ \hline
 & $0.25$  & $\pm 0.085 $ & $\langle \bfH_0 \rangle$ only\\
& &   $\pm 0.406$& $\langle \bfV_0 \rangle$ only\\ \hline
 & $1.0$ &  $\pm 0.217\, i$ & $\langle \bfH_0 \rangle$ only\\ 
& &   $\pm 0.437$ & $\langle \bfV_0 \rangle$ only\\ \hline
$128$ & $0.1$  & $\pm 0.342$ & $\langle \bfH_0 \rangle$ only\\ 
& &   $\pm1.276\, i$ & $\langle \bfV_0 \rangle$ only\\ \hline
\end{tabular}
\end{center}
\caption{Eigenvalues of $\boldsymbol{\mathcal{M}}$ for the MW$+$ forcing; $Re=1$.}\label{tab:evMWpM}
\end{table}

We then consider the full problem, as described in \S\,\ref{sec:analysis}(\ref{sec:full}), and compute the eigenvalues of $\boldsymbol{\mathcal{M}}$ by solving the system (\ref{eqn:indZEROFLUCT}) -- (\ref{eqn:contZEROFLUCT}); the elements of $\boldsymbol{\mathcal{M}}$ are contained in~\ref{app:matrices}. It appears that for all the values of $\mathcal{B}$ and $Rm$ investigated here, its eigenvectors are either purely magnetic or purely hydrodynamic, so that the mean magnetic and velocity fields do not grow at the same rate. The results for the eigenvalues and the nature of the corresponding eigenvectors are given in table~\ref{tab:evMWpM}.
For $Rm=16$, $\mathcal{B} = 0.25$ and $1$, the eigenvalue with the largest real part corresponds to a mean flow solution. This might be due to the fact that a non-zero mean flux introduces the anisotropy favourable to the development of an AKA-type instability. For $\mathcal{B}=0.25$, the eigenvalue corresponding to a mean magnetic field solution is pure imaginary, so this mode is marginally stable here, in contrast to the growing mode found for magnetic only solutions.
For $Rm=128$ and $\mathcal{B}=0.1$, the situation is reversed, as the eigenvalue with the largest real part corresponds to a mean magnetic field solution, possibly because higher values of $Rm$ (for fixed $Re$) favour magnetic field growth.

\subsection{Growth of $z$-dependent perturbations}

We now calculate the growth rate of $z$-dependent perturbations. We consider the artificial magnetic problem, for which we impose $\bfV = \textbf{0}$, as well as the full problem, in which magnetic and velocity perturbation equations are solved in concert, as in Hughes \& Proctor (2009a). For the latter we are able to determine the growth rate of both the magnetic and velocity eigenvectors by following the time evolution of the spatially averaged $\bfH(x,y,t)\, e^{p(k)t}$ and $\bfV(x,y,t) \,e^{p(k)t}$.

\begin{figure}
\begin{center}
\includegraphics[scale = 0.42]{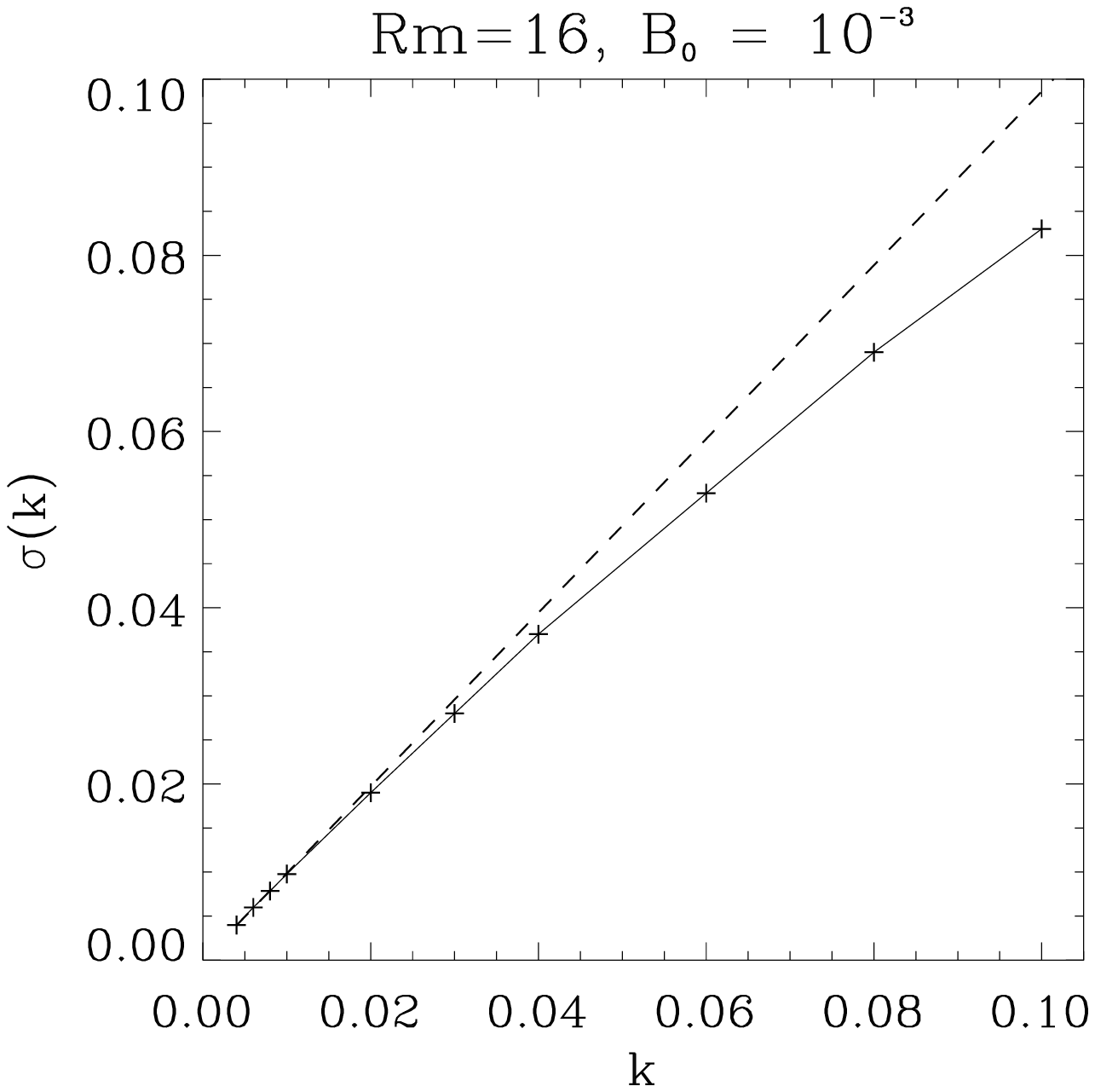}
\includegraphics[scale = 0.42]{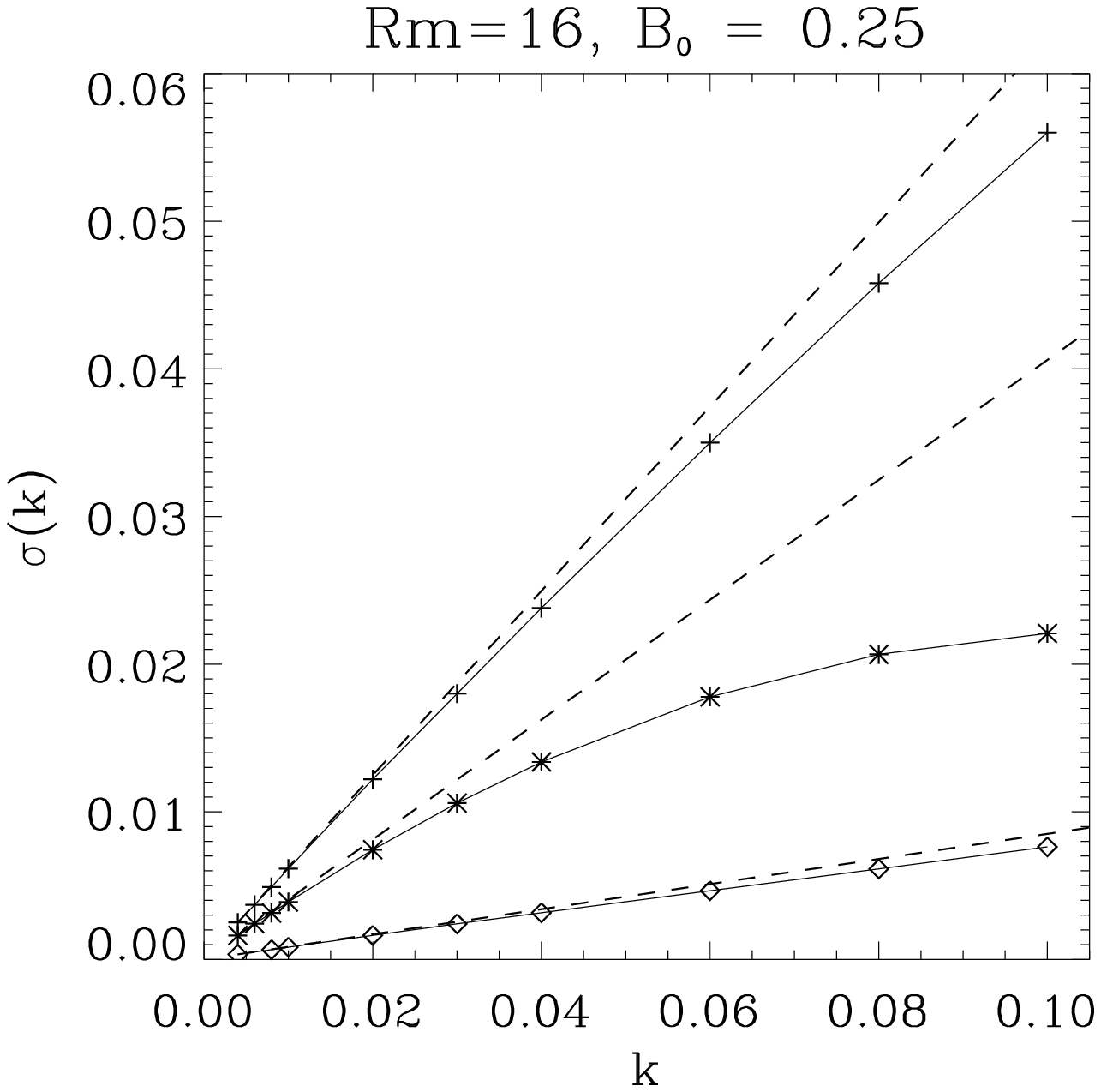}
\end{center}
\caption{\label{fig:sigMW1} Left: $\sigma(k)$ for $Rm = 16$, $\mathcal{B}=0.001$. The dashed line corresponds to $\lambda\, k$, where $\lambda$ is the positive eigenvalue of $\boldsymbol{\mathcal{A}}$.
Right: $\sigma(k)$ for the magnetic problem ($+$) and the full problem ($\star$, velocity mode, and $\diamond$, magnetic mode) for $Rm = 16$ and $\mathcal{B}=0.25$. The dashed lines corresponds to $\lambda\, k$, where $\lambda$ stands for the positive eigenvalue(s) of $\boldsymbol{\mathcal{A}}(\mathcal{B})$ (for the magnetic problem) and of $\boldsymbol{\mathcal{M}}$ (for the full problem).}
\end{figure}

\begin{figure}
\begin{center}
\includegraphics[scale = 0.42]{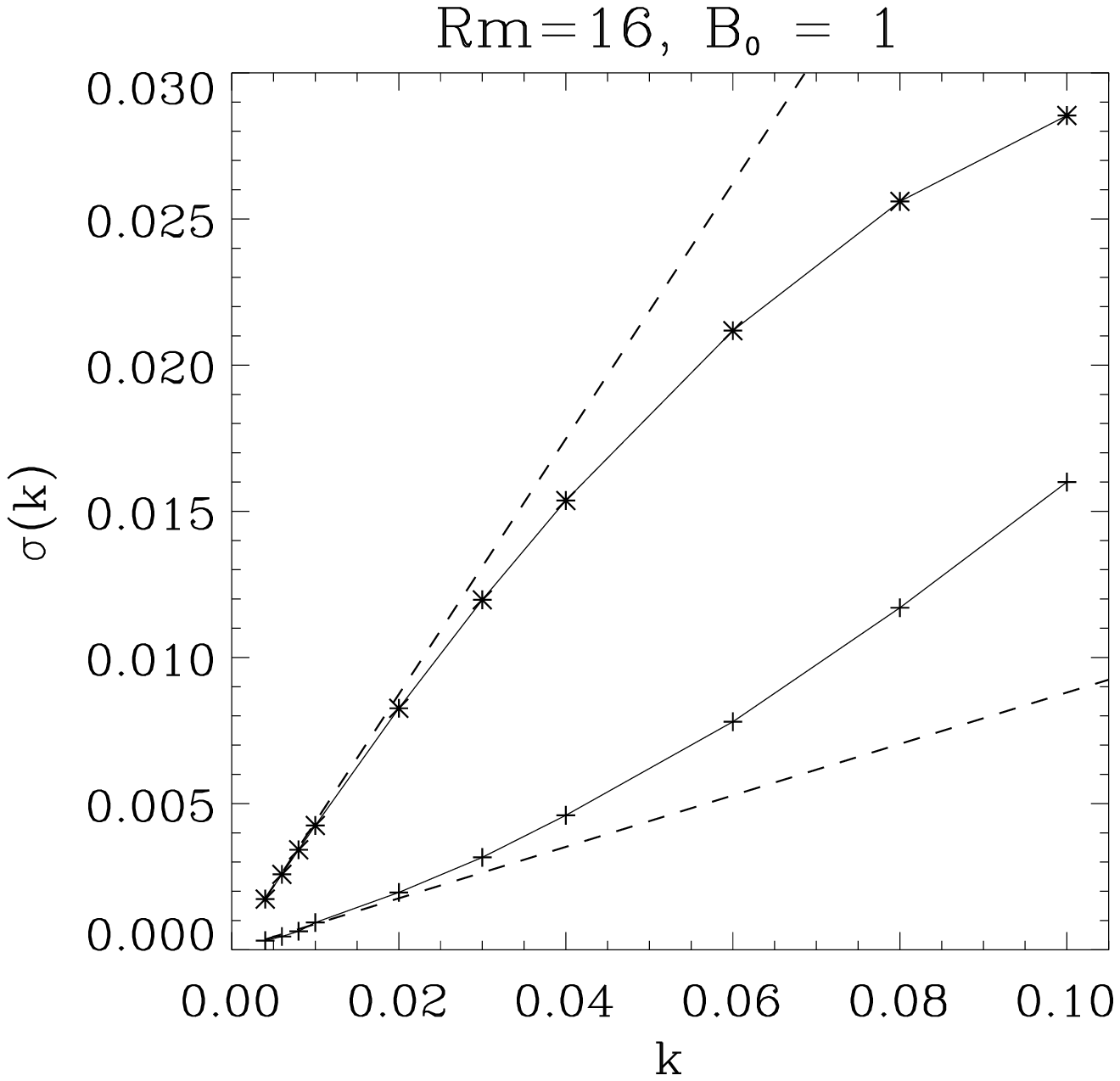}
\includegraphics[scale = 0.42]{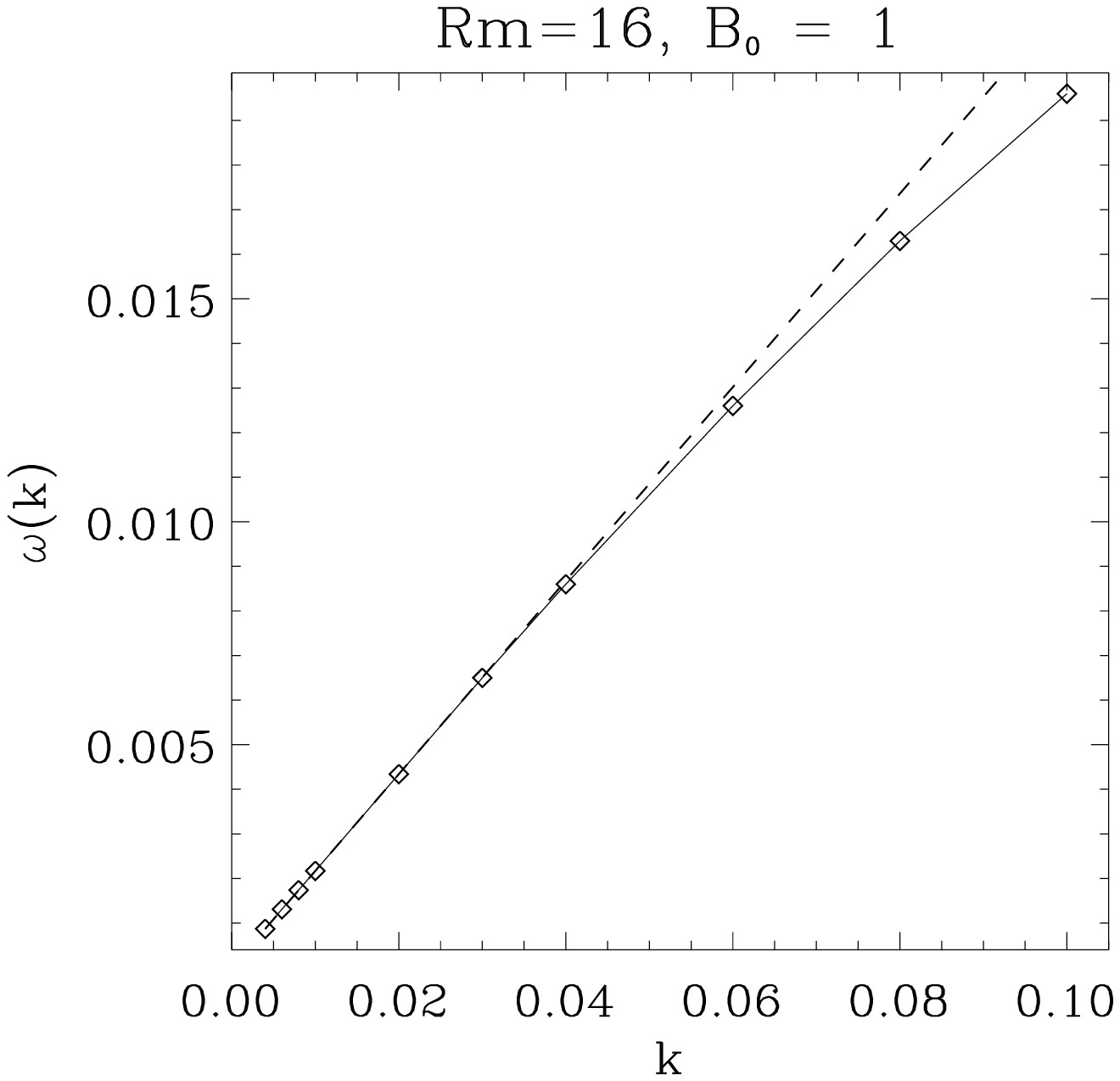}
\includegraphics[scale = 0.42]{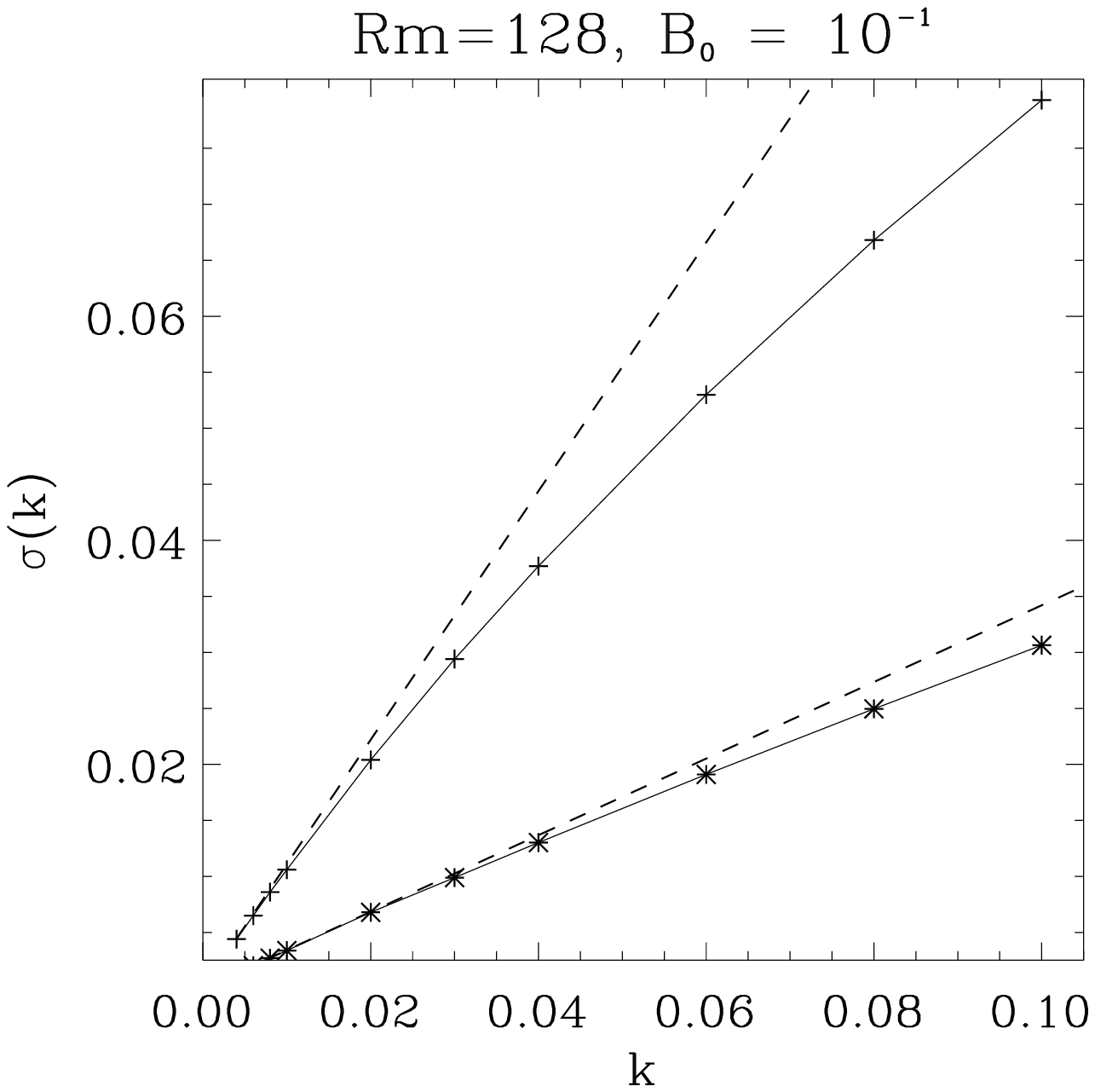}
\includegraphics[scale = 0.42]{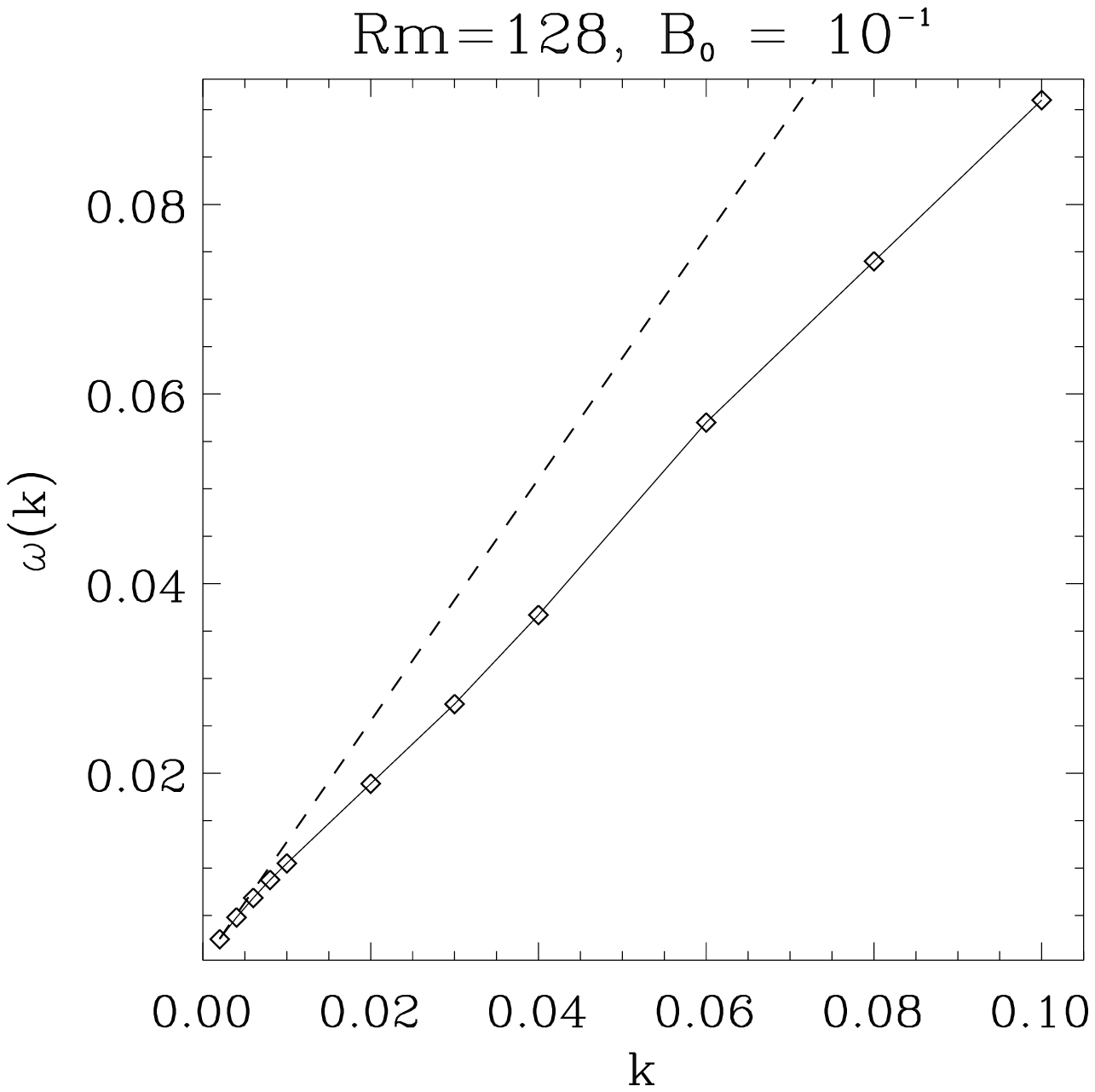}
\end{center}
\caption{\label{fig:sigMW2} 
Results for the MW$+$ forcing; $Rm=16$, $\mathcal{B} = 1.0$ (top plots); $Rm=128$, $\mathcal{B} = 0.1$ (bottom plots).
Left: $\sigma(k)$ for the magnetic problem ($+$) and the full problem ($\star$). The dashed lines corresponds to $Re[\lambda]\, k$, where $\lambda$ stands for the eigenvalue with the largest real part 
of $\boldsymbol{\mathcal{A}}(\mathcal{B})$ (for the magnetic problem) and of $\boldsymbol{\mathcal{M}}$ (for the full problem). Right: $\omega(k)$ for the full problem. The dashed lines corresponds to $Im[\lambda]\, k$, where $\lambda$ is the eigenvalue of $\boldsymbol{\mathcal{M}}$ corresponding to the marginal mode.}
\end{figure}

Figure~\ref{fig:sigMW1} shows $\sigma(k)$ for $Rm = 16$, with $\mathcal{B} = 0.001$ (left panel) and $0.25$ (right panel). The $+$ correspond to the fastest growing mode in the magnetic problem, while the $\star$ and $\diamond$ correspond to the dominant velocity and magnetic modes, respectively, in the full problem. Figure~\ref{fig:sigMW2} shows the results for  $Rm=16$ with $\mathcal{B} = 1$ and for $Rm=128$ with $\mathcal{B} = 0.1$. The left panels show $\sigma(k)$ for the fastest growing mode in the magnetic ($+$) and full ($\star$) problems. The right panels show $\omega(k)$ for the marginal mode in the full problem. In all cases, the dashed lines give the relevant linear approximation for low $k$ based on the eigenvalues of $\boldsymbol{\mathcal{A}}(\mathcal{B})$ or $\boldsymbol{\mathcal{M}}$. This shows that our results are in excellent agreement with the analysis presented in \S\,\ref{sec:analysis}. These results also demonstrate that in the cases investigated here, the growth rates of magnetic perturbations taken on their own are very different from those of the combined magnetic and velocity perturbations, which we believe to be the relevant situation to study. As shown by this particular example, this difference persists even in situations for which the relevant components of $\bfalpha^U$ and $\bfGamma^B$ vanish (see~\ref{app:matrices}).

\section{Conclusions}\label{sec:ccl}

In this paper we have investigated the instability of two-dimensional magnetohydrodynamic states to long-wavelength three-dimensional disturbances. As well as calculating the growth rates directly, we have used the methods pioneered by Roberts (1970) for the simpler kinematic dynamo, in order to derive a mean field theory for the linear growth of these perturbations, with coefficients that can be determined by imposing spatially uniform perturbations on the basic state. The calculated transport coefficients in the mean field theory give excellent agreement with the directly computed results.

The crucial feature of the mean field equations is that perturbations in the velocity and magnetic field are coupled, and that, in general, the growing eigenfunctions involve both sorts of field. Thus it is imperative that the induction and momentum equations are considered on the same footing. In contrast, earlier studies of the effects of equilibrated fields on the traditional $\alpha$-effect have used the induction equation without any coupling to the momentum equation, and this `passive' theory gives incorrect results in general.

Our results shed a new light on the general problem of the long-wavelength instabilities of saturated MHD states. In the past, the sole focus of interest has been the nature of the $\alpha$-effect, and its dependence on ambient magnetic field levels ($\alpha$-quenching). However, we can now see that there are situations in which the preferred unstable eigenmode may be dominated by the velocity field, and, furthermore, that there are cases for which magnetic and velocity perturbations grow faster than magnetic perturbations taken on their own. Thus even when strong (`catastrophic') quenching is predicted on the basis of the old ideas, we now see that there could be other mechanisms leading to a large-scale instability.

Here we have concentrated on 2D examples, which, although somewhat artificial, have the great advantage of allowing us to compare the predictions of mean field theory with the actual growth rate of the magnetic field, because we are able to isolate a single mode artificially --- something that would be much harder to achieve in 3D with the current computational resources. However, we emphasize that our main findings apply to the linear long-wavelength instability of any nonlinearly saturated MHD system, whether in two or three dimensions. There are however important differences between the 2D and 3D cases; a derivation of the formal mean field equations in the 3D case is given in~\ref{app:3D}, and a full treatment will be given in a subsequent paper (Courvoisier \textit{et al.}\ 2010).

\section*{Acknowledgements}
We are extremely grateful to Andrew Gilbert and Steve Tobias for many enlightening discussions. This work was supported by STFC and by a Royal Society Leverhulme Trust Senior Research Fellowship (DWH). Some of the work was performed during the 2009 workshop on Dynamo Theory at the Institut Henri Poincar\'e; we are very grateful to the organisers, Emmanuel Dormy, Stephan Fauve and Fran\c{c}ois P\'etr\'elis.

\appendix{The general case of a 3D basic state}\label{app:3D}

Here we derive the more general equations for a 3D basic state $\bfB(\bfx,t)$, $\bfU(\bfx,t)$, $P(\bfx, t)$, which we shall assume, for simplicity, to be spatially and temporally periodic. The calculations below take into account the possibility that $\langle \bfB \rangle \neq \textbf{0}$ and $\langle \bfU \rangle \neq \textbf{0}$. The basic state therefore can be the outcome of a saturated small-scale dynamo, or it can result from the presence of a mean flux, as in the 2D case. It can of course be a mixture of both.

Again, we are interested in the growth rate of long-wavelength perturbations, with wave vector $\bfk$, which we consider to be of the form
\begin{equation}
\left(\bfb(\bfx,t),\bfu(\bfx,t), p(\bfx,t) \right) =\left(\bfH(\bfx,t), \bfV(\bfx,t),\Pi(\bfx,t)\right)\,e^{i\bfk\cdot\bfx +p(\bfk)t}, \label{eqn:kansatz_2}
\end{equation}
where $\bfV$, $\bfH$ and $\Pi$ vary on the same spatial and temporal scales as the basic state, and $p(\bfk)$ is the $\bfk$-dependent growth rate of the perturbations. The wavelength of the perturbations, $2 \pi / |\bfk|$, is assumed to be long in comparison with the spatial periodicity of the basic state.
We now decompose the perturbations into mean and fluctuating parts, with the average defined by
\begin{equation}
\left(\overline \bfb , \overline \bfu , \overline p \right) = \left( \langle\bfH\rangle,\langle\bfV\rangle,\langle\Pi\rangle\right)\,e^{i\bfk\cdot\bfx +p(\bfk)t},
\label{}
\end{equation}
where $\langle \ \rangle$ denotes an average over the spatial and temporal periodicities of the basic state; $\langle\bfH\rangle$, etc.\ are thus constants. Our aim is to employ a mean field approach in order to determine the growth rate $p(\bfk)$ of modes with non-zero mean magnetic fields, i.e.\ modes for which $\langle \bfH \rangle \neq \textbf{0}$.

The evolution equations for the perturbations (cf.\ (\ref{eqn:indZPERT}) -- (\ref{eqn:contZPERT})) are
\begin{align}
\left(p+Rm^{-1}k^2\right) \bfH +\left( \partial_t - Rm^{-1}\nabla^2 \right) \bfH  &= \left(\nabla + i\bfk \right)\times \left( \bfV \times \bfB + \bfU \times \bfH\right)\nonumber \\
&+ 2iRm^{-1}\bfk \cdot \nabla \bfH,
\label{eqn:3D1} \\
\left(p+Re^{-1}k^2\right) \bfV +\left( \partial_t - Re^{-1}\nabla^2 \right) \bfV &= \left(\nabla + i\bfk \right)\cdot \left( \bfH\bfB + \bfB\bfH -\bfV\bfU - \bfU\bfV\right)\nonumber \\
 &-\nabla\Pi- i\bfk \Pi+ 2iRe^{-1}\bfk \cdot \nabla \bfV,
\label{eqn:3D2} \\
\nabla \cdot \bfH + i \bfk \cdot \bfH = 0, \quad
&\nabla \cdot \bfV + i \bfk \cdot \bfV = 0. \label{eqn:3D3}
\end{align}
As in the 2D analysis, we use $k = |\bfk|$ as a small parameter and expand $\bfH$, $\bfV$, $\Pi$ and the growth rate $p$ in powers of $k$, so that, for example, 
\begin{equation}
\bfH = \bfH_0+\bfH_1+ \ldots \bfH_n +\ldots,
\label{eqn:3D4}
\end{equation}
where $\bfH_n$ is of $n^{th}$ order in the components of $\bfk$. Each $\bfH_n$ can be decomposed into its mean and fluctuating parts, $\bfH_n = \langle \bfH_n \rangle + \bfh_n$; similarly, $\bfV_n = \langle \bfV_n \rangle + \bfv_n$.

At zeroth order, equations~(\ref{eqn:3D1}) -- (\ref{eqn:3D3}) become
\begin{eqnarray}
p_0\bfH_0 + \left(\partial_t - Rm^{-1}\nabla^2 \right) \bfH_0  = \nabla\times \left( \bfV_0 \times \bfB + \bfU \times \bfH_0\right),
\label{eqn:3D5} \\
p_0 \bfV_0 +\left( \partial_t - Re^{-1}\nabla^2 \right) \bfV_0 =-\nabla\Pi_0 + \nabla \cdot \left( \bfH_0\bfB + \bfB\bfH_0 -\bfV_0\bfU - \bfU\bfV_0\right),
\label{eqn:3D6} \\
\nabla \cdot \bfH_0 = 0, \quad
\nabla \cdot \bfV_0 = 0. \label{eqn:3D7}
\end{eqnarray}
On averaging these equations, we obtain
\begin{equation}
p_0 \langle \bfH_0 \rangle = p_0\langle \bfV_0 \rangle = 0.
\label{eqn:3D8}
\end{equation}
For non-zero mean field solutions we therefore need to take $p_0 = 0$. The corresponding equations for the fluctuations (cf.\ (\ref{eqn:indZEROFLUCT}) -- (\ref{eqn:contZEROFLUCT})) then read
\begin{align}
(\partial_t - Rm^{-1}\nabla^2 ) \bfh_0  &= 
\nabla\times \left( \bfv_0 \times \bfB + \bfU \times \bfh_0\right)+ \langle \bfH_0 \rangle \cdot \nabla\bfU -\langle \bfV_0 \rangle\cdot\nabla\bfB,
\label{eqn:3D9} \\
( \partial_t - Re^{-1}\nabla^2 ) \bfv_0 &=-\nabla\Pi_0 + \nabla \cdot \left( \bfh_0\bfB + \bfB\bfh_0 -\bfv_0\bfU - \bfU\bfv_0\right)\nonumber \\
&+ \langle \bfH_0 \rangle \cdot \nabla\bfB -\langle \bfV_0 \rangle\cdot\nabla\bfU,
\label{eqn:3D10} \\
\nabla \cdot \bfh_0 = 0, \quad
&\nabla \cdot \bfv_0 = 0. \label{eqn:3D11}
\end{align}
To first order, the equations for the mean variables (cf.\ (\ref{eqn:1stavU}) -- (\ref{eqn:1stavdivhv})) give
\begin{align}
&p_1 \langle \bfH_0 \rangle = i\bfk\times \langle \bfv_0 \times \bfB + \bfU \times \bfh_0 \rangle, \label{eqn:3D12} \\
&p_1 \langle \bfV_0 \rangle = -i\bfk \Pi_0 +i\bfk\cdot \langle \bfh_0 \bfB + \bfB\bfh_0 - \bfv_0\bfU -\bfU\bfv_0 \rangle, \label{eqn:3D13} \\
&\bfk \cdot \langle\bfH_0 \rangle= 0, \quad
\bfk \cdot \langle\bfV_0 \rangle= 0. \label{eqn:3D14} 
\end{align}
It can be seen from equations~(\ref{eqn:3D9}) -- (\ref{eqn:3D10}) that $\bfh_0$ and $\bfv_0$ are subject to a linear forcing by both $\langle \bfH_0 \rangle$ and $\langle \bfV_0 \rangle$. Therefore, expressions~(\ref{eqn:3D12}) and (\ref{eqn:3D13}) can be rewritten as
\begin{align}
&p_1 \langle \bfH_0 \rangle = i\bfk\times \left( \bfalpha^{B}\cdot\langle \bfH_0 \rangle + \bfalpha^{U}\cdot\langle \bfV_0 \rangle \right), \label{eqn:3D15} \\
&p_1 \langle \bfV_0 \rangle = -i\bfk \Pi_0 +i\bfk\cdot \left( \bfGamma^{B}\cdot\langle \bfH_0 \rangle + \bfGamma^{U}\cdot\langle \bfV_0 \rangle \right), \label{eqn:3D16} 
\end{align}
where the tensors $\bfalpha^{B}$, $\bfalpha^{U}$, $\bfGamma^{B}$ and $\bfGamma^{U}$ are constant and depend on the basic state $\bfU$, $\bfB$, and hence on the forcing $\bfF$ and on the Reynolds numbers. \\

These four tensors need to be evaluated for an accurate determination of the growth rate of the perturbations to first order in $k$. In theory, they can be calculated by solving equations~(\ref{eqn:3D12}) -- (\ref{eqn:3D14}). In practice however, these equations might have exponentially growing solutions independently of the forcing due to the mean magnetic and velocity fields (though if they are considered as long time solutions of an initial value problem this does not seem likely); we then may encounter the same problem as undermines the determination of the kinematic $\alpha$-effect in the case when a small-scale dynamo is present (Cattaneo \& Hughes 2009).

\appendix{Components of the matrices}\label{app:matrices}

In tables~\ref{tab:matAKA} and~\ref{tab:matMWp} we detail the components of the matrices whose eigenvalues are given in the main text, for the AKA forcing and the MW$+$ forcing respectively.

\begin{table}
\begin{center}
\small{\begin{tabular}{|c|c|}
\hline
$\mathcal{B}$ & Relevant matrices \\ \hline
$0$ &
$\boldsymbol{\mathcal{A}} = i\left[ \begin{array}{cc} 
-1.7667 & -1.7684  \\ 
\quad 1.7684 & \quad 1.7667   
\end{array}\right], \qquad
\boldsymbol{\mathcal{G}} = i\left[ \begin{array}{cc} 
\quad 0.033 & \quad 0.386  \\ 
-0.386 & -0.033   
\end{array}\right] $\\ \hline
$0.1$ &
$\boldsymbol{\mathcal{M}} = i\left[ \begin{array}{cccc} 
-1.311 & -1.297 & -0.376 & \quad 0.405 \\ 
\quad 1.310 & \quad 1.294 & \quad 0.486 & -0.295 \\ 
\quad 0.203 & \quad 0.236 & \quad 0.067 & \quad 0.296 \\ 
\quad 0.273 & \quad 0.229 & -0.305 & -0.038 
\end{array}\right]
$ \\ \hline
\end{tabular}}
\end{center}
\caption{Components of the matrices $\boldsymbol{\mathcal{A}}$, $\boldsymbol{\mathcal{G}}$ and $\boldsymbol{\mathcal{M}}$ for the AKA forcing and different values of $\mathcal{B}$; see section~\ref{sec:numAKA}.\label{tab:matAKA}}
\end{table}


\begin{table}
\small{\begin{center}
\begin{tabular}{|c|c|c|c|}
\hline
$Rm$ & $\mathcal{B}$ &  $\boldsymbol{\mathcal{M}}$ & $\boldsymbol{\mathcal{A}}(\mathcal{B})$\\ \hline

$16$ & $0.001$ &
$i\left[ \begin{array}{cccc} 
0 & 0.988 & 0 & 0 \\ 
-0.988 & 0 & 0 & 0 \\ 
0 & 0 & 0 & 0 \\ 
0 & 0 & 0 & 0 
\end{array}\right]
$
&
$i\left[ \begin{array}{cc} 
0 & 0.988  \\ 
-0.988 & 0   
\end{array}\right]$\\ \hline

 & $0.25$ &
$i\left[ \begin{array}{cccc} 
0 & 0.379 & 0 & 0 \\ 
-0.019 & 0 & 0 & 0 \\ 
0 & 0 & 0 & 0.815 \\ 
0 & 0 & -0.202 & 0 
\end{array}\right]
$
&
$i\left[ \begin{array}{cc} 
0 & 0.690  \\ 
-0.565 & 0   
\end{array}\right]$\\ \hline

 & $1.0$ &
$i\left[ \begin{array}{cccc} 
0 & 1.228 & 0 & 0 \\ 
0.038 & 0 & 0 & 0 \\ 
0 & 0 & 0 & 0.375 \\ 
0 & 0 & -0.509 & 0 
\end{array}\right]
$
&
$i\left[ \begin{array}{cc} 
0 & 0.053  \\ 
-0.145 & 0   
\end{array}\right]$\\ \hline

$128$ & $0.1$ &
$i\left[ \begin{array}{cccc} 
0 & 0.495 & 0 & 0 \\ 
-0.237 & 0 & 0 & 0 \\ 
0 & 0 & 0 & 1.429 \\ 
0 & 0 & 1.140 & 0 
\end{array}\right]
$
&
$i\left[ \begin{array}{cc} 
0 & 1.233  \\ 
-1.0 & 0   
\end{array}\right]$\\ \hline

\end{tabular}
\end{center}}
\caption{\label{tab:matMWp}Components of the matrices $\boldsymbol{\mathcal{M}}$ and $\boldsymbol{\mathcal{A}}(\mathcal{B})$ for the MW$+$ forcing and different values of $\mathcal{B}$; see section~\ref{sec:numMWp}.}
\end{table}

\end{document}